\documentclass[preprint,superscriptaddress,amsmath,amssymb,aps,eqsecnum,nofootinbib]{revtex4}
\usepackage{subfigure}
\usepackage{bm}
\usepackage{mathrsfs}

\usepackage{graphicx, color}

\usepackage{url}

\allowdisplaybreaks[2]

\def\cE{\mathcal{E}}

\def\cO{\mathcal{O}}

\def\mint{\int_{-\infty}^\infty\!\cdots\!\int_{-\infty}^\infty}
\def\l{\ell}

\newcommand{\be}{\begin{equation}}
\newcommand{\ee}{\end{equation}}
\newcommand{\ba}{\begin{aligned}}
\newcommand{\ea}{\end{aligned}}

\def\({\left(}
\def\){\right)}

\begin{document}
\title{Perturbative quasinormal mode frequencies}

\author{Yasuyuki Hatsuda}
\affiliation{Department of Physics, Rikkyo University, Toshima, Tokyo 171-8501, Japan}

\author{Masashi Kimura}
\affiliation{Department of Informatics and Electronics, Daiichi Institute of Technology, Tokyo 110-0005, Japan}
\affiliation{Department of Physics, Rikkyo University, Toshima, Tokyo 171-8501, Japan}
\date{\today}
\preprint{RUP-23-13}

\begin{abstract}
We often encounter a situation where linear wave equations around a black hole solution can be regarded as continuous deformations of simpler ones, or modifications from the general relativity case by continuous parameters.
We develop a general framework to compute high-order perturbative corrections to quasinormal mode frequencies in such deformed problems.
Our method has many applications, and allows to compute numerical values of the high-order corrections very accurately. For several examples, we perform this computation explicitly, and discuss analytic properties of the quasinormal mode frequencies for deformation parameters.
\end{abstract}

\maketitle

\tableofcontents

\renewcommand{\thefootnote}{\arabic{footnote}}
\setcounter{footnote}{0}
\setcounter{section}{0}

\section{Introduction}
Perturbation theory is one of the most powerful tools in physics.
We have a typical situation that a system cannot be solved analytically but its special limit can be.
Perturbation around the special limit provides us a good approximation method and more importantly a clue to get global information on the total system by combining with the analytic continuation in complex analysis or asymptotic analysis.
The application range of perturbation theory is extremely wide. It is important to clarify what we can learn about from perturbation theory.

In this work, we propose a systematic way to compute high-order perturbative corrections to quasinormal mode (QNM) frequencies of black holes.
QNMs are solutions to linearized field equations, which
satisfy purely ingoing (outgoing) boundary conditions at the horizon (infinity),
around a background black hole spacetime.
It is known that QNMs are related to the late time behavior of the field dynamics around black holes~\cite{Nakamura:1987zz, Kokkotas:1999bd, Nollert:1999ji, berti2009, konoplya2011, Ferrari:2007dd, 
Leaver:1986gd, Andersson:1996cm, Andersson:1995zk, Nollert:1992ifk, Berti:2006wq}.
In many cases, one can regard some parameters of black hole solutions as smooth deformation parameters of simpler black holes.
We apply perturbation theory for such deformation parameters.
Similar situations also happen if one considers possibilities of effective field theories or modified gravity theories beyond general relativity. 
Since such modification parameters are expected to be small, it is natural to expand physical quantities perturbatively. 
It is desirable to develop a general framework widely applicable for such cases.

There are two obstacles to achieve it.
One is that we cannot solve the QNM spectral problem analytically even in spherically symmetric black holes.
Therefore we have only numerical or semi-analytic eigenvalues and eigenfunctions in this simplest case.
The other point is more serious. In the QNM problem, a set of the eigenfunctions is not complete in the usual sense.
This means that we cannot apply the well-known formula in quantum mechanics to the computation of perturbative corrections to the QNM spectrum.
There is already an extended formula to compute perturbative corrections to QNM frequencies \cite{leung1997, leung1998, leung1999}.
However it is not clear for us how to use this formulation for our interested examples systematically and practically. For this reason, we revisit the similar problem in this work, and propose another way to get high-order perturbative corrections to QNM frequencies.

A possible resolution for this problem is simply to use numerical fittings.%
\footnote{There is another resolution. One can analytically continue eigenvalue problems from the real line to the complex domain. This is well-known as the complex scaling method for resonance problems in quantum mechanics.}
However it is hard to predict high-order corrections accurately in this way.
Recently, a smart way to compute perturbative quadratic order corrections to the QNM frequencies was proposed in \cite{cardoso2019, mcmanus2019}.
We are strongly motivated by these works. 
We extend them to more general setups.
Our approach is based on the first principle of perturbation theory.
We do not use any numerical fittings to determine the perturbative coefficients though we need numerical solutions at each order in perturbation. 
Our method is quite general and applicable to various situations with smoothly continuous deformations.
In fact, we give, for the first time, the high-precision perturbative expansion of the QNM frequencies around the extremal Reissner-Nordstr\"om black holes.
Combining a recently proposed method \cite{hatsuda2020}, our approach allows to compute numerical values of high-order corrections very accurately.
Once we get the high-order perturbative data, we can discuss analytic properties (convergence, singularity, analytic continuation, non-perturbative effect etc.) of the QNM frequencies in principle.

The organization is as follows. In Section~\ref{sec:general}, we start by explaining a general framework of our formulation. We illustrate our basic idea to compute high-order perturbative corrections systematically. In Section~\ref{sec:BW}, we present a technical way to perform the idea in Section~\ref{sec:general} explicitly. 
In Section~\ref{sec:examples}, we show various examples in which our method works well. 
We particularly use the method proposed in \cite{hatsuda2020}, but this is not only the possibility. For instance, we give another way in Sec.~\ref{app:series}. 
In Section~\ref{sec:outlook}, we consider possible future directions. In Appendix~\ref{app:ejidentities}, we give some remarks on the so-called parameterized black hole quasinormal mode approach. These are useful by combining the results in the main text. 

\section{General framework}\label{sec:general}
We first illustrate our idea.
In this section, we set up a problem, and explain a conceptual way to obtain perturbative series of QNM frequencies systematically. We will show a technical method to achieve it in the next section.
We expect that the problem proposed in this section is solved in many other ways developed in numerical computations of QNMs, such as Leaver's continued fraction method~\cite{leaver1985, konoplya2011}, the direct integration method~\cite{Chandrasekhar:1975zza, konoplya2011} or the pseudospectral method~\cite{Jansen:2017oag}.
In the following discussion, we do not need to assume that the effective potential 
in the master equation is given by an analytic function. 
However, if we apply our formalism to the Bender-Wu approach or Leaver's method as discussed in Secs.~\ref{sec:BW} and \ref{sec:examples}, the effective potential needs to be an analytic function.

We consider a perturbative deformation of a black hole in a certain theory.
We would like to know perturbative corrections to quasinormal mode frequencies for a small deformation parameter.
Our starting point is the following (radial) master equation:%
\footnote{Our idea is not restricted to this form. To make an explanation simpler, we assume it in this paper.}
\begin{equation}
\begin{aligned}
\( \frac{d^2}{dx^2}+\omega^2-V(x) \) \Phi(x)=0,
\end{aligned}
\label{eq:master}
\end{equation}
where $x$ is the tortoise coordinate whose range is $-\infty < x < \infty$
and the potential $V$ is defined in the same domain.
We assume that $V$ takes zero at $|x| \to \infty$.\footnote{
If the field has a mass term $\mu^2$, $V$ is constant at $x \to \infty$.
In that case, $\omega$ in Eq.~\eqref{eq:BC} should be changed into $\sqrt{\omega^2 -\mu^2}$
at $x \to \infty$. 
Then, we can still apply the same method. See subsection~\ref{subsec:massive}.
}
The tortoise coordinate $x$
is related to the radial variable $r$ as
\begin{equation}
\begin{aligned}
\frac{dx}{dr}=\frac{1}{f(r)}.
\end{aligned}
\label{eq:tortoise}
\end{equation}
where $f(r)$ is a function that has a zero at the event horizon $r=r_H$
and takes positive values outside the event horizon.\footnote{
When $f$ has another zero at $r = r_C $ in $r > r_H$, e.g., the case of Schwarzschild-de Sitter black hole, we focus on the region $r_H < r < r_C$.
}
Explicit forms of $f(r)$, of course, depend on problems.
The QNM boundary condition is then given by the purely outgoing/ingoing condition\footnote{
Because the wave function $\Phi$ with the QNM boundary condition
is divergent at $|x| \to \infty$ for ${\rm Im}(\omega) < 0$, one may think that 
the asymptotic form $\Phi(x) \sim e^{\pm i \omega x}$ is not sufficient to specify the boundary condition.
In fact, the QNM boundary condition is firstly defined in the domain ${\rm Im}(\omega) >0$,
then it corresponds to the decaying modes at $|x| \to \infty$, which is the fine-tuned modes.
By the analytic extension to the complex $\omega$ plane, we can define the QNM boundary condition for ${\rm Im}(\omega) < 0$.
}
\begin{equation}
\begin{aligned}
\Phi(x) \sim e^{\pm i \omega x} \quad (x \to \pm \infty).
\end{aligned}
\label{eq:BC}
\end{equation}

We assume that all the quantities in the master equation have smooth perturbative expansions in a parameter $\alpha$:
\begin{equation}
\begin{aligned}
V(x)=\sum_{k=0}^\infty \alpha^k V_k(x) ,\qquad
\omega^2=\sum_{k=0}^\infty \alpha^k \cE_k, \qquad
\Phi(x)=\sum_{k=0}^\infty \alpha^k \Phi_k(x).
\end{aligned}
\end{equation}
Typically, the parameter $\alpha$ appears as a deformation parameter of a black hole or of a modified theory. At this stage, we do not ask its physical origin for generality.
In general, the function $f(r)$ may also depends on $\alpha$. This dependence causes subtlety on our perturbative treatment. We will discuss this issue later.

Expanding $\omega$ as a series of $\alpha$, 
\begin{align}
\omega = \sum_{k=0}^{\infty} \alpha^k \omega_k,
\end{align}
the QNM boundary condition in Eq.~\eqref{eq:BC} can be written as
\begin{align}
\Phi &\sim 
 e^{\pm i \omega_0  x}
e^{\pm i (\alpha \omega_1 + \alpha^2 \omega_2 + \cdots) x}
\notag\\&=
 e^{\pm i \omega_0  x}
(1 + \alpha P^\pm_1 + \alpha^2 P^\pm_2 + \cdots),
\end{align}
where $P^\pm_1, P^\pm_2, \cdots$ are polynomials of $x$.
This implies that 
the QNM boundary condition for $\Phi_k$ is
\begin{align}
\Phi_k \sim  e^{\pm i \omega_0  x} \quad ( x \to \pm \infty).
\label{eq:BC-k}
\end{align}

We solve the master equation perturbatively in $\alpha$.
We start with the zeroth order, at which the eigen-equation is\footnote{
In our setup, by adding the factor $g$ introduced in Sec.~\ref{sec:BW} and taking the analytic continuation, the problem reduces to the eigenvalue problem for one-dimensional bound states whose eigenvalues are not degenerate.
Thus, the zeroth order spectra are not degenerate.
}
\begin{equation}
\begin{aligned}
\( \frac{d^2}{dx^2}+\cE_0-V_0(x) \) \Phi_0(x)=0,\qquad \cE_0=\omega_0^2.
\end{aligned}
\label{eq:zeroth}
\end{equation}
Note that $\cE_0$ denotes the zeroth order eigenvalue in perturbation of $\alpha$, not the fundamental mode eigenvalue.
Typically, the zeroth order equation is the master equation for spherically symmetric black holes,
but our formalism is not restricted to this specific situation.
At each order, we solve the differential equation by requiring proper boundary conditions, and then get the perturbative corrections to the eigenvalues.

We first solve the zeroth order equation \eqref{eq:zeroth} by imposing the ordinary QNM boundary condition:
\begin{equation}
\begin{aligned}
\Phi_0(x) \sim e^{\pm i \omega_0 x} \quad ( x \to \pm \infty).
\end{aligned}
\end{equation}
There are many techniques to solve Eq.~\eqref{eq:zeroth} numerically.
To go to the next order, we need 
the zeroth order eigenfunction $\Phi_0(x)$ with the eigenvalue $\cE_0$. 
\footnote{This point is quite different from the textbook-like method in quantum mechanics, in which one needs all the eigenvalues and the eigenfunctions at the zeroth order to use them as a basis of Hilbert space.}
Though, in this work, we will use a method recently proposed in \cite{hatsuda2020},
we stress that our idea should work for many other techniques.

Once we obtain the eigenvalue and the eigenfunction at the zeroth order,
we can proceed to the first order equation.
The equation we should solve is
\begin{equation}
\begin{aligned}
\( \frac{d^2}{dx^2}+\cE_0-V_0(x) \) \Phi_1(x)=(V_1(x)-\cE_1)\Phi_0(x).
\end{aligned}
\label{eq:first}
\end{equation}
We regard this equation as the inhomogeneous differential equation for $\Phi_1(x)$ with the unknown constant $\cE_1$, while $\Phi_0(x)$ and $\cE_0$ are known.
For the function $\Phi_1(x)$, we impose the same QNM boundary condition for $\Phi_0(x)$:
\begin{equation}
\begin{aligned}
\Phi_1(x) \sim e^{\pm i \omega_0 x} \quad ( x \to \pm \infty),
\end{aligned}
\label{eq:1stBC}
\end{equation}
as explained in Eq.~\eqref{eq:BC-k}.
As shown in the next section, this inhomogeneous equation is also solved by the same method as the zeroth order equation.
Therefore, we get $\cE_1$ and $\Phi_1(x)$ at least numerically.
We note that $\cE_1$ is uniquely determined for a given zeroth order $\cE_0$.\footnote{
Assuming that Eq.~\eqref{eq:first} has two solutions with the appropriate QNM boundary condition
$\Phi_1^{\rm (i)}$ and $\Phi_1^{\rm (ii)}$ whose 
eigenvalues are $\cE_1^{\rm (i)}$ and $\cE_1^{\rm (ii)}$, respectively, 
the deviation $\Delta \Phi_1 := \Phi_1^{\rm (i)} - \Phi_1^{\rm (ii)}$ satisfies
an equation 
$( d^2/dx^2 +\cE_0-V_0) \Delta \Phi_1  =-\Delta \cE_1 \Phi_0$,
where $\Delta \cE_1 :=\cE_1^{\rm (i)} - \cE_1^{\rm (ii)}$.
This equation is same as Eq.~\eqref{eq:first} with a vanishing correction term $V_1 = 0$.
Thus, the only possible solution is $\Delta \Phi_1 \propto \Phi_0$ with $\Delta \cE_1 = 0$.
This implies $\cE_1^{\rm (i)} = \cE_1^{\rm (ii)}$.
}

The computations at higher orders are similar. We regard the $k$-th order equation
\begin{equation}
\begin{aligned}
\( \frac{d^2}{dx^2}+\cE_0-V_0(x) \) \Phi_k(x)=\sum_{\ell=1}^{k} (V_{\ell}(x)-\cE_{\ell})\Phi_{k-\ell}(x).
\end{aligned}
\label{eq:k-th}
\end{equation}
as the inhomogeneous equation for $\cE_k$ and $\Phi_k(x)$ with the known $\cE_j$ and $\Phi_j(x)$ ($0\leq j \leq k-1$).
We solve it under the boundary condition in Eq.~\eqref{eq:BC-k}.
We repeat this computation as many times as possible.

If the function $f(r)$ depends on the perturbative parameter $\alpha$, there is a subtle point.
In this case, we also expand $f(r)$ in $\alpha$. This gives a perturbative relation between $r$ and $x$ via the relation in Eq.~\eqref{eq:tortoise}.
Schematically, we have
\begin{equation}
\begin{aligned}
x=x(r,\alpha)=\sum_{k=0}^\infty \alpha^k x_{k}(r),
\end{aligned}
\label{eq:tortoise-2}
\end{equation}
where $x_{k}(r)$ are functions of $r$. On the other hand, we can inverse this relation by
\begin{equation}
\begin{aligned}
r=r(x,\alpha)=\sum_{k=0}^\infty \alpha^k r_k(x).
\end{aligned}
\label{eq:tortoise-3}
\end{equation}
There is an ambiguity which variable, $r$ or $x$, is fundamental in the perturbative expansion.
In this paper, we regard $x$ as a fundamental variable, and use Eq.~\eqref{eq:tortoise-3} to eliminate $r$ to expand the potential perturbatively.
This is because boundary conditions in terms of $x$ seem to be more natural.

There is a caveat when we
apply our framework to a specific system 
and calculate the QNM frequencies by numerical calculations.
Our framework is introduced based on the form of the master equation in Eq.~\eqref{eq:master} which is written by the tortoise  coordinate $x$.
However, in many cases, it is difficult to explicitly write the tortoise coordinate $x$ as a function of $r$ 
and also the master equation as a function of $x$.
This implies that imposing the boundary condition at each order $\Phi_k \sim  e^{\pm i \omega_0  x}$
is not a trivial task in a concrete example. 
In that case, the technique to rewrite the master equation used in~\cite{cardoso2019} might be useful.
When the function $f$ has a zero at $r = r_H$, 
and it is close to $1-r_H/r$,
we can write $f$ as
\begin{align}
f = \left(1 - \frac{r_H}{r}\right) Z(r;\alpha), 
\end{align}
where $Z(r;\alpha)$ is a function of $r$ which contains the small parameter $\alpha$.
Choosing $r_H$ and $\alpha$ as the fundamental parameters, we can write the master equation Eq.~\eqref{eq:master} in the form
\begin{align}
\left(1 - \frac{r_H}{r}\right) \frac{d}{dr} \left( \left(1 - \frac{r_H}{r}\right) \frac{d\phi}{dr}  \right)  + (\tilde{\omega}^2 - \tilde{V}) \phi = 0,
\end{align}
where $\phi = \sqrt{Z}\Phi$, 
$\tilde{\omega}$ is a rescaled frequency and $\tilde{V}$ is the effective potential which depends on $\alpha$~\cite{cardoso2019}.\footnote{
The explicit forms of 
$\tilde{\omega}$ and $\tilde{V}$ can be seen in Appendix.~B in~\cite{cardoso2019}.
}
Regarding this equation as the basic master equation, 
we can easily apply our framework to this system 
because the tortoise coordinate in this system 
is explicitly written as 
$r + r_H \ln (1-r_H/r)$.
We should note that 
we do not need to care about this point
as far as we use the Bender-Wu approach introduced in the next section
because the calculation is carried out around potential peak region.

Finally note that our formulation is easily extended to multi-parameter perturbations.
If one wants to consider a two-parameter perturbation:
\begin{equation}
\begin{aligned}
V(x;\alpha,\beta)=V_0(x)+\sum_{k=1}^\infty (\alpha^k V_{k}^{\alpha}(x)+\beta^k V_{k}^{\beta}(x)),
\end{aligned}
\label{eq:multi-para-pot}
\end{equation}
then the square of the frequency should receive the following perturbative corrections \cite{cardoso2019, mcmanus2019}:
\begin{equation}
\begin{aligned}
\omega^2&=\cE_0+\alpha \cE_1^{(1,0)}+\beta \cE_1^{(0,1)}+\alpha^2 \cE_2^{(2,0)}+\alpha\beta \cE_2^{(1,1)}+\beta^2 \cE_2^{(0,2)}+\cdots \\
&=\cE_0+\sum_{k=1}^\infty \sum_{\ell=0}^{k} \alpha^\ell \beta^{k-\ell} \cE_{k}^{(\ell, k-\ell)} .
\end{aligned}
\label{eq:multi-para}
\end{equation}
To fix the coefficients $\cE_k^{(\ell, k-\ell)}$, we can choose various combinations of $(\alpha,\beta)$.
For instance, to fix the second order corrections $\cE_2^{(2,0)}$, $\cE_2^{(1,1)}$ and $\cE_2^{(0,2)}$, it is sufficient to consider three particular slices: $(\alpha,\beta)\to (\alpha,0),(\alpha, \alpha), (0,\alpha)$, in which the problem is reduced to the one-parameter problem.
We will return to this issue in Section~\ref{sec:examples}.

\section{Technical remark: the Bender-Wu approach}\label{sec:BW}

In the previous section, we proposed a general idea to compute the perturbative corrections $\cE_k$ systematically.
The main problem is of course how we solve the differential equation \eqref{eq:k-th} for our interested QNM problems.
In this section, we see that this is done by the so-called Bender-Wu approach \cite{bender1969} that is recently extended to the QNM computation in \cite{hatsuda2020, eniceicu2020}, based on \cite{blome1984, ferrari1984, Ferrari:1984zz}. 
The main advantage of this approach is that it is widely applicable to many models, as in the WKB approach \cite{mashhoon1983, schutz1985}.
The Bender-Wu approach itself also highly depends on perturbation theory. 
Since we need eigenfunctions as well as eigenvalues, we review the Bender-Wu approach for our problem.
We follow the notation in \cite{sulejmanpasic2018} as much as possible.

\subsection{Leading order solution}
Let us solve the zeroth order equation \eqref{eq:zeroth}.
We first introduce a formal parameter $g$ by hand,
\begin{equation}
\begin{aligned}
\( -g^4 \frac{d^2}{dx^2}+\cE_0-V_0(x) \) \Phi_0(x)=0,\qquad \cE_0=\omega_0^2.
\end{aligned}
\label{eq:zeroth-BW}
\end{equation}
It is clear to see that $g^2$ plays the role of a Planck parameter. Setting $g=e^{\pi i/4}$, the original equation \eqref{eq:zeroth} is reproduced.%
\footnote{Note that there is another possibility: $g=e^{-\pi i/4}$. This ambiguity reflects the fact that the QNM frequencies have two branches for the real part \cite{hatsuda2020}.}
The basic idea is the following. We first consider the eigenvalue problem for $g \in \mathbb{R}$.
In this case, we have the Schr\"odinger-type equation with the \textit{inverted} potential $-V_0(x)$, which admit bound states, and we can apply the standard perturbative method in quantum mechanics near the minimum of $-V_0(x)$. The important observation in \cite{hatsuda2020} is that the boundary conditions for bound states and QNMs are simply related by the analytic continuation of $g$. This implies that if we know the bound state energy for $g \in \mathbb{R}$, we can obtain the QNM eigenvalue by the analytic continuation $g=e^{\pi i/4}$.

Let $\bar{x}$ be the value of $x$ at which $-V_0(x)$ takes the minimal value.
We expand the inverted potential $-V_0(x)$ around $x=\bar{x}$:
\begin{equation}
\begin{aligned}
-V_0(x)=V_{00}+\sum_{j=2}^\infty V_{0j} (x-\bar{x})^j.
\end{aligned}
\end{equation}
We introduce a new variable by $x-\bar{x}=gq$.
This change means that as $g$ decreases, we zoom in on the neighborhood of the minimum at $x=\bar{x}$.
Then Eq.~\eqref{eq:zeroth-BW} leads to
\begin{equation}
\begin{aligned}
\( -\frac{1}{2} \frac{d^2}{dq^2}+\frac{1}{2}\Omega^2 q^2+v_0(q)-\epsilon_0 \)\psi_0(q)= 0,
\end{aligned}
\label{eq:BW-0}
\end{equation}
where $\Omega:=\sqrt{V_{02}}$, $\epsilon_0:=-(\cE_0+V_{00})/(2g^2)$ and 
\begin{equation}
\begin{aligned}
v_0(q)=\frac{1}{2g^2} \sum_{j=3}^\infty V_{0j} (gq)^j=\sum_{j=1}^\infty g^{j} v_{0j} q^{j+2}, \quad v_{0j}:=\frac{V_{0,j+2}}{2}.
\end{aligned}
\end{equation}
We denoted $\psi_0(q)=\Phi_0(\bar{x}+gq)$ to avoid confusion.
In this picture, the Planck constant is unity, and $g$ now plays the role of a coupling constant in the potential.

We solve Eq.~\eqref{eq:BW-0} perturbatively in $g$ order by order.
At the leading order, we can regard it as the harmonic oscillator with frequency $\Omega$.
To eliminate the exponential factor of the eigenfunction, we rescale $\psi_0(q)=e^{-\Omega q^2/2}u_0(q)$:
\begin{equation}
\begin{aligned}
-\frac{1}{2}u_0''(q)+\Omega q u_0'(q)+\( \frac{\Omega}{2}+v_0(q)-\epsilon_0\) u_0(q)=0.
\end{aligned}
\label{eq:diff-u0-0}
\end{equation}
We have the following expansions:
\begin{equation}
\begin{aligned}
u_0(q)=\sum_{n=0}^\infty g^n u_{0n}(q),\qquad
\epsilon_0=\sum_{n=0}^\infty g^n  \epsilon_{0n}.
\end{aligned}
\label{eq:pert-0}
\end{equation}
Plugging these expansions into Eq.~\eqref{eq:BW-0}, we get
\begin{equation}
\begin{aligned}
-\frac{1}{2}u_{0n}''+\Omega q u_{0n}'+\frac{\Omega}{2} u_{0n}+\sum_{j=1}^n v_{0j} q^{j+2} u_{0,n-j}-\sum_{j=0}^n \epsilon_{0j} u_{0,n-j}=0.
\end{aligned}
\end{equation}

Let us focus on the ground state for simplicity. The ground state corresponds to the lowest (or fundamental) overtone mode in the QNM problem.
For $n=0$, we have the trivial solution $u_{00}(q)=1$ and $\epsilon_{00}=\Omega/2$.
Using it, we get
\begin{equation}
\begin{aligned}
-\frac{1}{2}u_{0n}''+\Omega q u_{0n}'+\sum_{j=1}^n (v_{0j} q^{j+2}-\epsilon_{0j}) u_{0,n-j}=0,\quad n \geq 1.
\end{aligned}
\label{eq:diff-u0}
\end{equation}
The very important fact is that $u_{0n}$ is a \textit{polynomial of $q$ whose degree is at most $3n$} \cite{bender1969, sulejmanpasic2018}:
\begin{equation}
\begin{aligned}
u_{0n}=\sum_{m=1}^{3n} A_{0n}^m q^m, \quad n \geq 1.
\end{aligned}
\label{eq:u_0n}
\end{equation}
As shown in \cite{bender1969}, the differential equation \eqref{eq:diff-u0} determines all the coefficients $A_{0n}^m$ and $\epsilon_{0n}$ recursively.
This is what the \textit{Mathematica} program in \cite{sulejmanpasic2018} is doing.
One has to keep in mind that the above result is valid only for the ground state. For the excited states, we need to modify it slightly.
See \cite{sulejmanpasic2018} for these cases.

We finally want to set $g=e^{\pi i/4}$ in the perturbative series. However, in general, the formal power series in Eq.~\eqref{eq:pert-0} are not convergent for any $g \ne 0$.
The substitution of $g=e^{\pi i/4}$ merely gives a meaningless answer.
To avoid it, one needs to truncate all the high-order corrections beyond a certain optimal order or to use summation methods. Note that the former turns out to be equivalent to the WKB series in the literature \cite{mashhoon1983, schutz1985}.
We use the latter, called the Borel summation method, to decode a meaningful result for finite $g$ from formal divergent series.\footnote{An alternative way is to use Pad\'e approximants \cite{matyjasek2017, konoplya2019, matyjasek2019}.} 
The conclusion in \cite{hatsuda2020} is that the Borel summation of Eq.~\eqref{eq:pert-0} correctly reproduces the QNM frequencies.
We emphasize that the above method allows us to construct not only the eigenvalue $\cE_0$ but also the eigenfunction $\psi_0(q)$. In summary, for the ground state, we have
\begin{equation}
\begin{aligned}
\cE_0&=-V_{00}-2g^2\sum_{n=0}^\infty g^n  \epsilon_{0n},\\
\psi_0(q)&=e^{-\Omega q^2/2} \sum_{n=0}^\infty g^n u_{0n}(q),\qquad
u_{0n}(q)=\sum_{m=1}^{3n} A_{0n}^m q^m,
\end{aligned}
\end{equation}
where $\epsilon_{00}=\Omega/2$ and $u_{00}(q)=1$.

\subsection{First order correction}

Let us proceed to the first order correction.
We need to solve
\begin{equation}
\begin{aligned}
\( -g^4 \frac{d^2}{dx^2}+\cE_0-V_0(x) \) \Phi_1(x)=( V_1(x)-\cE_1) \Phi_0(x).
\end{aligned}
\label{eq:first-order}
\end{equation}
Note that we already know the zeroth order eigenfunction $\Phi_0(x)$ and eigenvalue $\cE_0$ in the previous subsection.
As in the computation above, we can rewrite it as
\begin{equation}
\begin{aligned}
\( -\frac{1}{2} \frac{d^2}{dq^2}+\frac{1}{2}\Omega^2 q^2+v_0(q)-\epsilon_0 \)\psi_1(q)=\frac{V_1(x)-\cE_1}{2g^2}\psi_0(q),
\end{aligned}
\end{equation}
We also expand $-V_1(x)$ around $x=\bar{x}$ as
\begin{equation}
\begin{aligned}
-V_1(x)
=\sum_{j=0}^\infty V_{1j} (gq)^j.
\end{aligned}
\end{equation}
Note that $x=\bar{x}$ does not extremize $V_1(x)$ in general.
As mentioned in the previous subsection, we have to impose the same boundary conditions for $\psi_0(q)$ and $\psi_1(q)$.
Therefore we set $\psi_1(q)=e^{-\Omega q^2/2} u_1(q)$ as well as $\psi_0(q)=e^{-\Omega q^2/2} u_0(q)$, and get
\begin{equation}
\begin{aligned}
-\frac{1}{2}u_1''+\Omega q u_1'+\( \frac{\Omega}{2}+v_0-\epsilon_0\) u_1
+\(\frac{V_{11}}{2g}q+v_1-\epsilon_1 \) u_0=0,
\end{aligned}
\label{eq:u1-diff}
\end{equation}
where
\begin{equation}
\begin{aligned}
\epsilon_1&:=-\frac{\cE_1+V_{10}}{2g^2} ,\\
v_1(q)&:=\frac{1}{2g^2} \sum_{j=2}^\infty V_{1j}(gq)^j=\sum_{j=0}^\infty g^j v_{1j} q^{j+2}, \qquad v_{1j}=\frac{V_{1,j+2}}{2}.
\end{aligned}
\end{equation}
We use the zeroth order perturbative solution in Eq.~\eqref{eq:pert-0}.
From the consistency at the orders $1/g^2$ and $1/g$, we should take
\begin{equation}
\begin{aligned}
u_1(q)=-\frac{V_{11}}{2\Omega g}q+\sum_{n=0}^\infty g^n u_{1n}(q),\qquad
\epsilon_1=\sum_{n=0}^\infty g^n \epsilon_{1n}.
\end{aligned}
\end{equation}
It is observed that for the ground state, $u_{1n}(q)$ is a polynomial of at most degree $3n+4$.
After putting an ansatz for the polynomial $u_{1n}(q)$, we can determine all the coefficients of $u_{1n}(q)$ and $\epsilon_{1n}$ from the perturbative equations.
The remaining computation is the same as the zeroth order one.
By performing the Borel summation of $\epsilon_1$, we obtain the first correction $\cE_1$.

\subsection{On higher order corrections}
The computations for higher orders are straightforward.
At the $k$-th order, we have
\begin{equation}
\begin{aligned}
\( -g^4 \frac{d^2}{dx^2}+\cE_0-V_0(x) \) \Phi_k(x)=\sum_{\ell=1}^{k}( V_{\ell}(x)-\cE_{\ell}) \Phi_{k-\ell}(x).
\end{aligned}
\end{equation}
It leads to
\begin{equation}
\begin{aligned}
-\frac{1}{2}u_k''+\Omega q u_k'+\( \frac{\Omega}{2}+v_0-\epsilon_0\) u_k+\sum_{\ell=1}^k \(\frac{V_{\ell 1}}{2g}q+v_\ell-\epsilon_\ell \) u_{k-\ell}=0,
\end{aligned}
\end{equation}
where $\Phi_k(x)=e^{-\Omega q^2/2} u_k(q)$ and 
\begin{equation}
\begin{aligned}
\epsilon_\ell:=-\frac{\cE_\ell+V_{\ell 0}}{2g^2},\qquad
v_\ell(q):=\frac{1}{2g^2} \sum_{j=2}^\infty V_{\ell j} (gq)^j.
\end{aligned}
\end{equation}
We observe that the ground state solution in general behaves as
\begin{equation}
\begin{aligned}
u_k(q)&=\frac{u_{k,-k}(q)}{g^k}+\cdots=\sum_{n=-k}^\infty g^n u_{kn}(q), \\
\epsilon_k&=\frac{\epsilon_{k,-2}}{g^2}+\cdots=\sum_{n=-1}^\infty g^{2n} \epsilon_{k,2n},
\end{aligned}
\end{equation}
where $u_{kn}(q)$ is a polynomial of at most degree $3n+4k$.
Under this assumption, we can easily compute $\epsilon_k$ perturbatively in $g$.

\section{Examples}\label{sec:examples}
In this section, we apply our formalism to various examples.

\subsection{A toy model: the Rosen-Morse potential}
We demonstrate that the idea in Section~\ref{sec:general} actually works in the QNM problem for a simple exactly solvable toy model.
What we consider is the so-called Rosen-Morse potential, which is regarded as an integrable deformation of the P\"oschl-Teller potential.
The Rosen-Morse potential was studied in the context of the quasinormal modes in massive scalar perturbations \cite{ohashi2004}.
We revisit the same model to validate our framework.
This model is given by
\begin{equation}
\begin{aligned}
&\( \frac{d^2}{dx^2}+\omega^2-V_\text{RM}(x) \) \phi(x)=0,\\
&V_\text{RM}(x)=\frac{1}{2\cosh^2 x}+\mu^2 \frac{1+\tanh x}{2}.
\end{aligned}
\label{eq:RM}
\end{equation}
where $\mu$ is a deformation parameter. If $\mu=0$, the potential reduces to the well-known P\"oschl-Teller potential.
The Rosen-Morse potential in Eq.~\eqref{eq:RM} for $\mu \ne 0$ is very similar to the potential for the spherically symmetric black hole in the massive scalar perturbation~\cite{ohashi2004}. We will see it in the next subsection.
We treat this system as a perturbation in the parameter $\mu$.

We first show that this system is in fact exactly solvable. To do so, we perform a change of variables and a transformation  of the wave function by
\begin{equation}
\begin{aligned}
z=\frac{1}{2}(1+\tanh x), \qquad \phi(x)=z^{-i\omega/2}(1-z)^{-i\sqrt{\omega^2-\mu^2}/2} y(z).
\end{aligned}
\end{equation}
Then, the new function $y(z)$ satisfies the standard hypergeometric equation:
\begin{equation}
\begin{aligned}
z(1-z)y''(z)+[c-(a+b+1)z]y'(z)-aby(z)=0,
\end{aligned}
\end{equation}
where
\begin{equation}
\begin{aligned}
a&=\frac{1}{2}-\frac{i}{2}(\omega+\sqrt{\omega^2-\mu^2}+1),  \\
b&=\frac{1}{2}-\frac{i}{2}(\omega+\sqrt{\omega^2-\mu^2}-1), \\
c&=1-i\omega. 
\end{aligned}
\end{equation}
For a given $\mu$, we impose the QNM-like boundary condition:
\begin{equation}
\begin{aligned}
\lim_{x \to -\infty} \phi(x) \sim e^{-i\omega x},\qquad
\lim_{x \to +\infty} \phi(x) \sim e^{+i\sqrt{\omega^2-\mu^2}\, x}, 
\end{aligned}
\label{eq:RM-BC0}
\end{equation}
where we have to choose a branch of the square root so that $\sqrt{z^2}=z$ for $z \in \mathbb{C}$ in order to match the boundary condition for $\mu=0$. 
In terms of $y(z)$, this boundary condition is translated into the regularity condition both at $z=0, 1$ simultaneously.
The regular solution at $z=0$ is given by the Gauss hypergeometric function
\begin{equation}
\begin{aligned}
y(z)=F(a,b;c;z).
\end{aligned}
\end{equation}
Using the well-known analytic connection formula of the hypergeometric function:
\begin{equation}
\begin{aligned}
&F(a,b,c;z)=\frac{\Gamma(c)\Gamma(c-a-b)}{\Gamma(c-a)\Gamma(c-b)} F(a,b,a+b-c+1;1-z)\\
&\qquad\qquad+\frac{\Gamma(c)\Gamma(a+b-c)}{\Gamma(a)\Gamma(b)} (1-z)^{c-a-b} F(c-a,c-b,c-a-b+1;1-z),
\end{aligned}
\end{equation}
the regularity condition at $z=1$ requires
\begin{equation}
\begin{aligned}
\frac{1}{\Gamma(a)\Gamma(b)}=0.
\end{aligned}
\end{equation}
Therefore we obtain $a=-n$ or $b=-n$ for $n=0,1,2,\dots$. This condition leads to the following exact spectrum:
\begin{equation}
\begin{aligned}
\omega^{(n, \pm)}=\pm \( \frac{1}{2}+\mu^2 \frac{1}{4(2n^2+2n+1)}\) -i\( n+\frac{1}{2} -\mu^2 \frac{2n+1}{4(2n^2+2n+1)} \).
\end{aligned}
\end{equation}
We have two symmetric branches of  the spectra.
The exact eigenfunction is also given by
\begin{equation}
\begin{aligned}
\phi^{(n, \pm)}(x)=\( \frac{1+\tanh x}{2} \)^{-i\omega^{(n, \pm)}} \( \frac{1-\tanh x}{2} \)^{-i\sqrt{\omega^{(n, \pm)2}-\mu^2}/2}\\
\times F\(-n,-n\mp i;1-i\omega^{(n, \pm)};\frac{1+\tanh x}{2} \)
\end{aligned}
\label{eq:RM-eigenfunction}
\end{equation}
Note that for a non-negative integer $n$, the hypergeometric function in this equation is a polynomial of degree $n$.
For simplicity, we consider the case of $b=-n$, and abbreviate the upper index in these expressions. For the lowest overtone number $n=0$, we have
\begin{equation}
\begin{aligned}
\omega&=
\frac{1-i}{2}+\mu^2 \frac{1+i}{4}, \\
\phi(x)&=\( \frac{1+\tanh x}{2} \)^{-i\omega/2} \( \frac{1-\tanh x}{2} \)^{-i\sqrt{\omega^2-\mu^2}/2}.
\end{aligned}
\end{equation}
In the small $\mu$ limit, we have
\begin{equation}
\begin{aligned}
\omega^2&=\cE_0+\mu^2\cE_1+\mu^4\cE_2=-\frac{i}{2}+\frac{\mu^2}{2}+\frac{i\mu^4}{8}, \\
\phi(x)&=\phi_0(x)+\mu^2 \phi_1(x)+\mu^4 \phi_2(x)+\cO(\mu^6),
\end{aligned}
\label{eq:RM-pert-1}
\end{equation}
where
\begin{equation}
\begin{aligned}
\phi_0(x)&=\( \frac{1}{2\cosh x} \)^{-i\omega_0} ,\\
\phi_1(x)&=\frac{1-i}{4}x \( \frac{1}{2\cosh x} \)^{-i\omega_0},\\
\phi_2(x)&=-\frac{i}{16}x^2 \( \frac{1}{2\cosh x} \)^{-i\omega_0},
\end{aligned}
\label{eq:RM-pert-2}
\end{equation}
and $\omega_0=(1-i)/2$.
These functions satisfy the same boundary condition:
\begin{equation}
\begin{aligned}
\lim_{x \to -\infty} \phi_k(x) \sim e^{-i\omega_0 x},\qquad
\lim_{x \to +\infty} \phi_k(x) \sim e^{+i\omega_0 x}, \qquad k=0,1,2,\dots.
\end{aligned}
\label{eq:RM-BC}
\end{equation}
Note that this boundary condition is slightly different from the true QNM boundary condition in Eq.~\eqref{eq:RM-BC0}, but after resumming the perturbative series it is reproduced correctly.

Now we confirm this result from perturbation theory.
We consider the perturbation in $\mu^2$:
\begin{equation}
\begin{aligned}
V_\text{RM}(x)&=V_0(x)+\mu^2 V_1(x), \\
V_0(x)&=\frac{1}{2\cosh^2 x},\qquad V_1(x)=\frac{1+\tanh x}{2}.
\end{aligned}
\end{equation}
At the lowest order, we of course obtain the P\"oschl-Teller potential:
\begin{equation}
\begin{aligned}
\( \frac{d^2}{dx^2}+\cE_0-V_0(x) \)\phi_0(x)=0,
\end{aligned}
\end{equation}
Its eigenvalue and the eigenfunction for the fundamental QNM are exactly given by the zeroth order in Eqs.~\eqref{eq:RM-pert-1} and \eqref{eq:RM-pert-2}.
We can confirm them by using the Bender-Wu approach in the previous section. We perturbatively solve Eq.~\eqref{eq:diff-u0-0} or \eqref{eq:diff-u0} for
\begin{equation}
\begin{aligned}
\Omega=\frac{1}{\sqrt{2}},\qquad v_0(q)=\frac{1}{2g^2}\biggl( -\frac{1}{2\cosh^2 (g q)}+\frac{1}{2}-\frac{(g q)^2}{2} \biggr).
\end{aligned}
\end{equation}
By putting the ansatz in Eqs.~\eqref{eq:pert-0} and \eqref{eq:u_0n}, we find the following perturbative expansions:
\begin{equation}
\begin{aligned}
\epsilon_0=\frac{1}{2 \sqrt{2}}-\frac{g^2}{4}+\frac{g^4}{8 \sqrt{2}}-\frac{g^8}{64 \sqrt{2}}+\frac{g^{12}}{256 \sqrt{2}}-\frac{5 g^{16}}{4096 \sqrt{2}}+\frac{7 g^{20}}{16384 \sqrt{2}}-\frac{21 g^{24}}{131072 \sqrt{2}}\\
+\frac{33 g^{28}}{524288 \sqrt{2}} -\frac{429 g^{32}}{16777216 \sqrt{2}}+\frac{715 g^{36}}{67108864 \sqrt{2}}-\frac{2431 g^{40}}{536870912 \sqrt{2}}+\cO(g^{44}),
\end{aligned}
\label{eq:epsilon0-RM}
\end{equation}
and
\begin{equation}
\begin{aligned}
u_0(q)&=1+g^2 \left(\frac{q^2}{4}+\frac{q^4}{12 \sqrt{2}}\right)+g^4 \left(-\frac{q^2}{8 \sqrt{2}}-\frac{q^4}{96}-\frac{q^6}{720 \sqrt{2}}+\frac{q^8}{576}\right)\\
&\quad+g^6 \left(-\frac{q^4}{96 \sqrt{2}}-\frac{11 q^6}{5760}+\frac{13 q^8}{40320 \sqrt{2}}-\frac{17 q^{10}}{34560}+\frac{q^{12}}{20736 \sqrt{2}}\right)+\cO(g^8).
\end{aligned}
\end{equation}
It is relatively easy to push high order computations. We performed it up to $\cO(g^{240})$.
Note that the perturbative series of $\epsilon_0$ is precisely reproduced by the exact result in \cite{berti2009},
\begin{equation}
\begin{aligned}
\epsilon_0=-\frac{g^2}{4}+\frac{1}{2}\sqrt{\frac{1}{2}+\frac{g^4}{4}}.
\end{aligned}
\label{eq:zerothe0}
\end{equation}
We also observe that the perturbative series of $u_0(q)$ is generated by the following analytic function:\footnote{Note that this analytic function behaves as $e^{\frac{q^2}{2\sqrt{2}}}$ in the large $|q|$ regime. This behavior is needed to reproduce the correct boundary condition of the original function $\phi_0(x)$, as seen in Eq.~\eqref{eq:zerothphi0}.}
\begin{equation}
\begin{aligned}
u_0(q)=e^{\frac{q^2}{2\sqrt{2}}} \biggl( \frac{1}{\cosh (gq)} \biggr)^{\frac{2\epsilon_0}{g^2}}.
\end{aligned}
\label{eq:zerothu0}
\end{equation}
Now we substitute $g=e^{\pi i/4}$ into Eqs.~\eqref{eq:zerothe0} and \eqref{eq:zerothu0}. Then we find
\begin{equation}
\begin{aligned}
\cE_0&=\frac{1}{2}-2g^2 \epsilon_0=-\frac{i}{2},\\
\phi_0(x)&=e^{-\frac{q^2}{2\sqrt{2}}}u_0(q)
\propto \left(\frac{1}{2\cosh x}\right)^{-i\omega_0}.
\end{aligned}
\label{eq:zerothphi0}
\end{equation}
These coincide with the exact results in Eqs.~\eqref{eq:RM-pert-1} and \eqref{eq:RM-pert-2}.
Of course, the Borel summation or the Pad\'e approximant of the perturbative series in Eq.~\eqref{eq:epsilon0-RM} also gives a good approximate eigenvalue.

At the first and the second orders, we have
\begin{equation}
\begin{aligned}
&\( \frac{d^2}{dx^2}+\cE_0-V_0(x) \)\phi_1(x)+(\cE_1-V_1(x))\phi_0(x)=0,\\
&\( \frac{d^2}{dx^2}+\cE_0-V_0(x) \)\phi_2(x)+(\cE_1-V_1(x))\phi_1(x)+\cE_2 \phi_0(x)=0.
\end{aligned}
\end{equation}
We would like to solve these inhomogeneous equations under the boundary condition in Eq.~\eqref{eq:RM-BC}.
Instead, it is sufficient to confirm that the functions in Eqs.~\eqref{eq:RM-pert-1} and \eqref{eq:RM-pert-2} satisfy these differential equations.
One can immediately check it.

These corrections are also reproduced by the Bender-Wu approach.
At the first order, we solve Eq.~\eqref{eq:u1-diff}. After some computations, we find
\begin{equation}
\begin{aligned}
\epsilon_1&=0,\\
u_1(q)&=\frac{q}{2\sqrt{2}g}+g\left( \frac{q}{4}+\frac{q^3}{8\sqrt{2}}+\frac{q^5}{48}\right)\\
&\quad+g^3\left(\frac{q}{8 \sqrt{2}}+\frac{q^3}{32}+\frac{q^5}{64 \sqrt{2}}-\frac{q^7}{2880}+\frac{q^9}{1152 \sqrt{2}}\right)+\cO(g^5).
\end{aligned}
\end{equation}
In this case, it is very likely that the first order correction $\epsilon_1$ does not receive any perturbative corrections. We confirmed it up to $\cO(g^{240})$. Therefore we have
\begin{equation}
\begin{aligned}
\cE_1=-V_{10}-2g^2 \epsilon_1=\frac{1}{2}.
\end{aligned}
\end{equation}
Similarly, at the second order, we find
\begin{equation}
\begin{aligned}
\epsilon_2=-\frac{1}{16 g^2}-\frac{1}{8 \sqrt{2}}-\frac{g^2}{16}-\frac{g^4}{32 \sqrt{2}}+\frac{g^8}{256 \sqrt{2}}-\frac{g^{12}}{1024 \sqrt{2}}+\frac{5 g^{16}}{16384 \sqrt{2}}-\frac{7 g^{20}}{65536 \sqrt{2}}\\
+\frac{21 g^{24}}{524288 \sqrt{2}}-\frac{33 g^{28}}{2097152 \sqrt{2}}+\frac{429 g^{32}}{67108864 \sqrt{2}}-\frac{715 g^{36}}{268435456 \sqrt{2}}
+\cO(g^{40}).
\end{aligned}
\end{equation}
and
\begin{equation}
\begin{aligned}
u_2(q)&=\frac{q^2}{16 g^2}+\left(\frac{q^2}{8 \sqrt{2}}+\frac{q^4}{64}+\frac{q^6}{192 \sqrt{2}}\right)\\
&\quad+g^2 \left(\frac{q^2}{16}+\frac{3 q^4}{128 \sqrt{2}}+\frac{7 q^6}{1536}-\frac{q^8}{11520 \sqrt{2}}+\frac{q^{10}}{9216}\right)+\cO(g^4).
\end{aligned}
\end{equation}
We observe that the second order correction $\epsilon_2$ are related to the zeroth order correction $\epsilon_0$ by
\begin{equation}
\begin{aligned}
\epsilon_2=-\frac{1}{16g^2}-\frac{g^2}{8}-\frac{\epsilon_0}{4}.
\end{aligned}
\end{equation}
Using this guess and setting $g=e^{\pi i/4}$, we finally get
\begin{equation}
\begin{aligned}
\cE_2=2g^2\epsilon_2=\frac{1+g^4-\cE_0}{4}=\frac{i}{8}.
\end{aligned}
\end{equation}
Our perturbative computation in the Bender-Wu approach implies $\cE_{k\geq 3}=0$ for any $g$. All of these results are consistent with the exact result.

For higher overtone modes, since the hypergeometric function in Eq.~\eqref{eq:RM-eigenfunction} does not change the asymptotic behavior of the solution,
the same structure holds.

\subsection{Massive scalar perturbations}\label{subsec:massive}
The simplest example in black hole problems is a massive scalar perturbation of the Schwarzschild geometry.
The functions in the master equation are given by
\begin{equation}
\begin{aligned}
f(r)=1-\frac{2M}{r},\qquad V(x)=f(r)\( \frac{\l(\l+1)}{r^2}+\frac{2M}{r^3}+\mu^2 \).
\end{aligned}
\label{eq:massive-funcs}
\end{equation}
As in the Rosen-Morse potential,
we regard the scalar mass square $\mu^2$ as a deformation parameter: $\alpha=\mu^2$. 
Note that the function $f(r)$ does not receive any correction. 
The explicit relation between $r$ and $x$ is given by
\begin{equation}
\begin{aligned}
x=r+2M \log \left( \frac{r}{2M}-1 \right).
\end{aligned}
\end{equation}
We regard $r$ as a function of $x$.
The unperturbed system is just the massless scalar case:
\begin{equation}
\begin{aligned}
V_0(x)=f(r)\( \frac{\l(\l+1)}{r^2}+\frac{2M}{r^3} \).
\end{aligned}
\end{equation}
The correction in the potential is
\begin{equation}
\begin{aligned}
V_1(x)=f(r),\qquad V_{k \geq 2}(x)=0.
\end{aligned}
\end{equation}
The QNM frequency receives the perturbative corrections in $\mu^2$. To keep the generality of $M$, we write the perturbative series as the dimensionless form
\begin{equation}
\begin{aligned}
M \omega=\sum_{k=0}^\infty (M \mu)^{2k} w_k,
\end{aligned}
\label{eq:massive-pert}
\end{equation}
where the correction coefficients $w_k$ does not depend on $M$.
Our task is to compute $w_k$ order by order.
We can apply the method in Section~\ref{sec:BW}. 

Let us briefly see the boundary condition. In the case of Eq.~\eqref{eq:massive-funcs}, the total boundary condition for the QNM is
\begin{equation}
\begin{aligned}
\lim_{x \to -\infty} \Phi(r)\sim e^{-i\omega x},\qquad
\lim_{x \to +\infty} \Phi(r)\sim e^{+i \sqrt{\omega^2 -\mu^2}\, x},
\end{aligned}
\end{equation}
If $\mu$ is small, the boundary condition at infinity is expanded as
\begin{equation}
\begin{aligned}
e^{+i \sqrt{\omega^2 -\mu^2}\, x}=e^{+i\omega x}\(1-\frac{ix}{2\omega}\mu^2 -\frac{(i+\omega x)x}{8\omega^3}\mu^4+\cO(\mu^6) \).
\end{aligned}
\end{equation}
This is indeed consistent with our requirement in Eq.~\eqref{eq:BC-k}. 

To show an explicit result, we focus on the cases of $\l=2, 3$.%
\footnote{As explained in \cite{hatsuda2020}, the Bender-Wu approach works well for larger $\ell$. This is why we consider $\ell=2,3$ rather than $\ell=0,1$.
It is desirable to solve Eq.~\eqref{eq:k-th} in other approaches.} 
It is sufficient for us to compute the coefficients in Eq.~\eqref{eq:massive-pert} for the case of $M=1$ actually.
The zeroth order frequency for the lowest overtone number%
\footnote{The reader should not confuse the subscript index here with the overtone number.} is well-known: 
\begin{equation}
\begin{aligned}
w_0^{\l=2}=0.4836438722 - 0.0967587760i,\quad
w_0^{\l=3}=0.6753662325 - 0.0964996277i.
\end{aligned}
\end{equation}
We have computed the numerical values of the perturbative coefficients $w_k$ up to $k=40$.
The first six values are shown in Table~\ref{tab:massive}.
In this table, we showed stable digits in our numerical computations. 
The leading and next-to-leading corrections are consistent with the early results in \cite{cardoso2019, mcmanus2019}.

\begin{table}[tp]
\caption{The first six perturbative corrections to the fundamental QNM frequency in Eq.~\eqref{eq:massive-pert} with $\l=2, 3$ in the massive scalar perturbation.}
\begin{center}
\begin{tabular}{c@{\hspace{10pt}}c@{\hspace{10pt}}c}
\hline
$k$ & $w_k^{\l=2}$ &  $4^k w_k^{\l=3}$ \\
\hline
$0$ &  $0.4836438722 - 0.0967587760i$ &  $0.6753662325 - 0.0964996277i$      \vspace{-5pt}\\
$1$ &  $0.3156326579 + 0.1081551348i$  & $0.9437297621 + 0.2278771948i$    \vspace{-5pt}\\
$2$ &  $0.03541170393 + 0.02620890155i$ & $0.2263735226 + 0.1075217988i$  \vspace{-5pt}\\
$3$ &  $0.01199156679 + 0.02204684913i$  & $0.2085153094 + 0.1986390780i$   \vspace{-5pt}\\
$4$ &  $0.00092115819 + 0.02209374509i$  &  $0.2333370885 + 0.4509679860i$  \vspace{-5pt}\\
$5$ &  $-0.01001596605 + 0.02211024342i$ &  $0.1500437709 + 1.0963976002i$     \vspace{-5pt}\\
$6$ &  $-0.02390151862 + 0.01898789685i$ &  $-0.580414699 + 2.681826119i$    \\
\hline
\end{tabular}
\end{center}
\label{tab:massive}
\end{table}%

\begin{figure}[tbp]
\begin{center}
\includegraphics[width=0.45\linewidth]{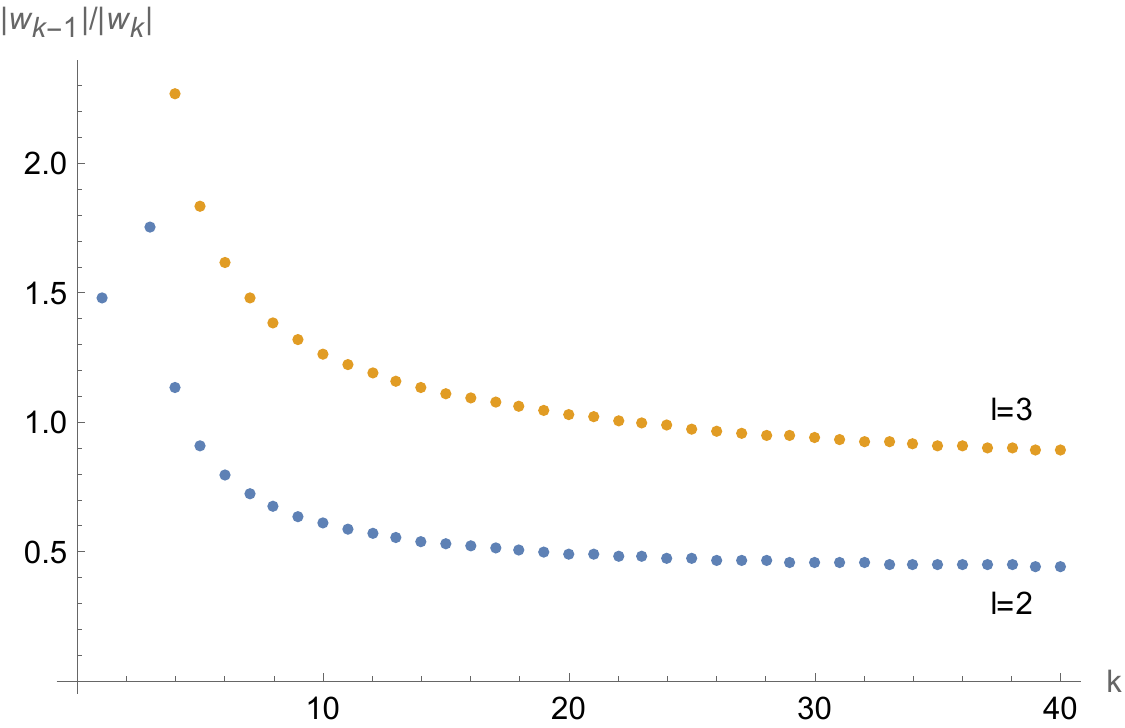}
\caption{To see whether the perturbative series \eqref{eq:massive-pert} is convergent or not, we plot the ratio $|w_{k-1}|/|w_k|$ for $1\leq k \leq 40$. It looks to converge to a finite value.}
\label{fig:ratio-w}
\end{center}
\end{figure}

What do we learn about from these perturbative data?
The most basic question would be whether the perturbative series in Eq.~\eqref{eq:massive-pert} is convergent or not.
To see it, we show the behavior of the ratio $w_{k-1}/w_{k}$ up to $k=40$ in figure~\ref{fig:ratio-w}.
The ratio seems to converge to a finite value, but the convergence is slow.
Using basic knowledge of complex analysis, we can estimate the radius of convergence in a different way.
The radius of convergence is determined by the nearest singular point from the origin.
In our framework, we have only the finite number of $w_k$. We would like to decode the singularity structure from these data. Probably the best tool to do so is \textit{Pad\'e approximants}.

\begin{figure}[tb]
\begin{center}
  \begin{minipage}[b]{0.45\linewidth}
    \centering
    \includegraphics[width=0.95\linewidth]{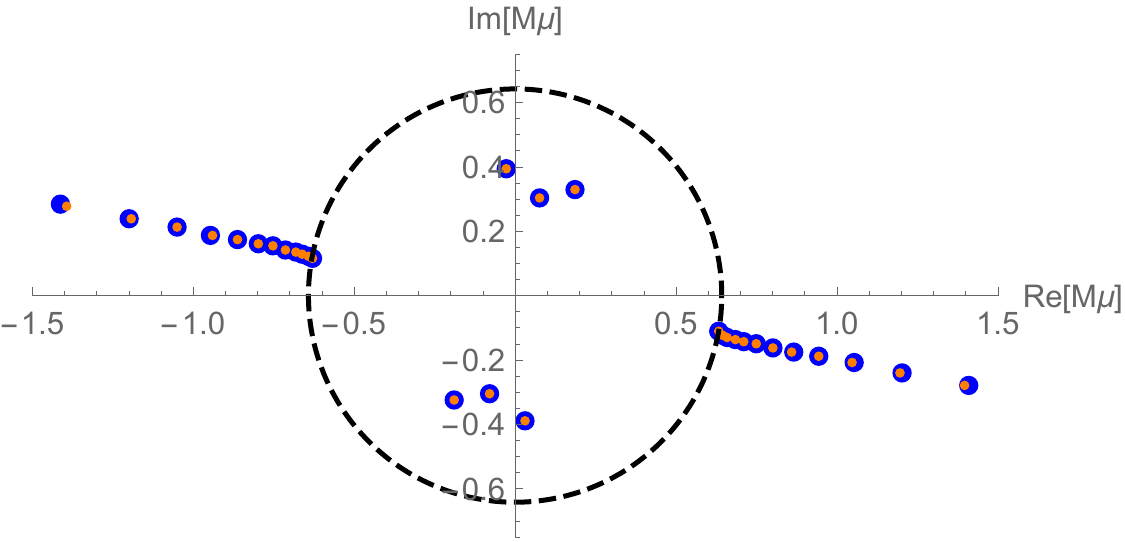}
  \end{minipage} \hspace{1cm}
  \begin{minipage}[b]{0.45\linewidth}
    \centering
    \includegraphics[width=0.95\linewidth]{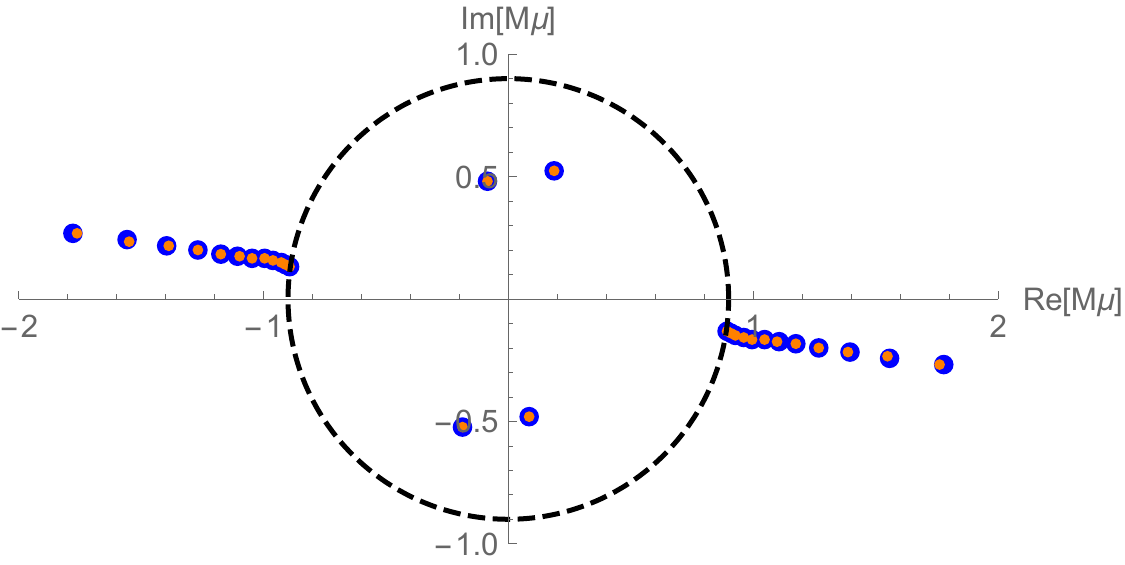}
  \end{minipage} 
\end{center}
  \caption{The singularity structure of the $[40/40]$ Pad\'e approximant of Eq.~\eqref{eq:massive-pert} for $\ell=2$ (Left) and $\ell=3$ (Right) in the complex $M\mu$-plane. We show its zeros by the blue points and poles by the orange points. The dashed curve is a conjectural convergence circle of the perturbative series in Eq.~\eqref{eq:massive-pert}. Note that the zeros and the poles inside the circle disappear when the degrees of the Pad\'e approximant are varied. These are artifacts for the $[40/40]$ Pad\'e approximant.}
  \label{fig:singularity}
\end{figure}

Pad\'e approximants tell us the analytic structure of a given power series.
In particular, it gives us information on sigularity structure on the original function. See Appendix C in \cite{hatsuda2021a}, for instance.
Since we have the perturbative data of Eq.~\eqref{eq:massive-pert} up to $(M\mu)^{80}$, we can construct its diagonal Pad\'e approximant $M\omega^{[40/40]}$. We read off the zeros and the poles of this approximant. The results are illustrated in Figure~\ref{fig:singularity}. These figures imply that the perturbative series in Eq.~\eqref{eq:massive-pert} is likely a convergent series.
One can estimate its radius of convergence by computing the distance to the nearest singular point.
In this computation, one has to be care about ``false'' singular points of Pad\'e approximants.
These singular points disappear if orders of Pad\'e approximants are changed. These are artifacts in the approximant, while the ``true'' singular points are stable for Pad\'e orders. In Figure~\ref{fig:singularity}, we observe that the black dashed circle is expected to be the convergence circle. The estimation of the radius of convergence $R$ for Eq.~\eqref{eq:massive-pert} in the complex $M \mu$-plane is approximately given by
\begin{equation}
\begin{aligned}
R_\text{fund.}^{\ell=2}\approx 0.643, \qquad R_\text{fund.}^{\ell=3} \approx 0.900. 
\end{aligned}
\end{equation}
We do not a clear physical meaning of this radius so far. It would be interesting to understand it.

\begin{figure}[tb]
\begin{center}
  \begin{minipage}[b]{0.45\linewidth}
    \centering
    \includegraphics[width=0.95\linewidth]{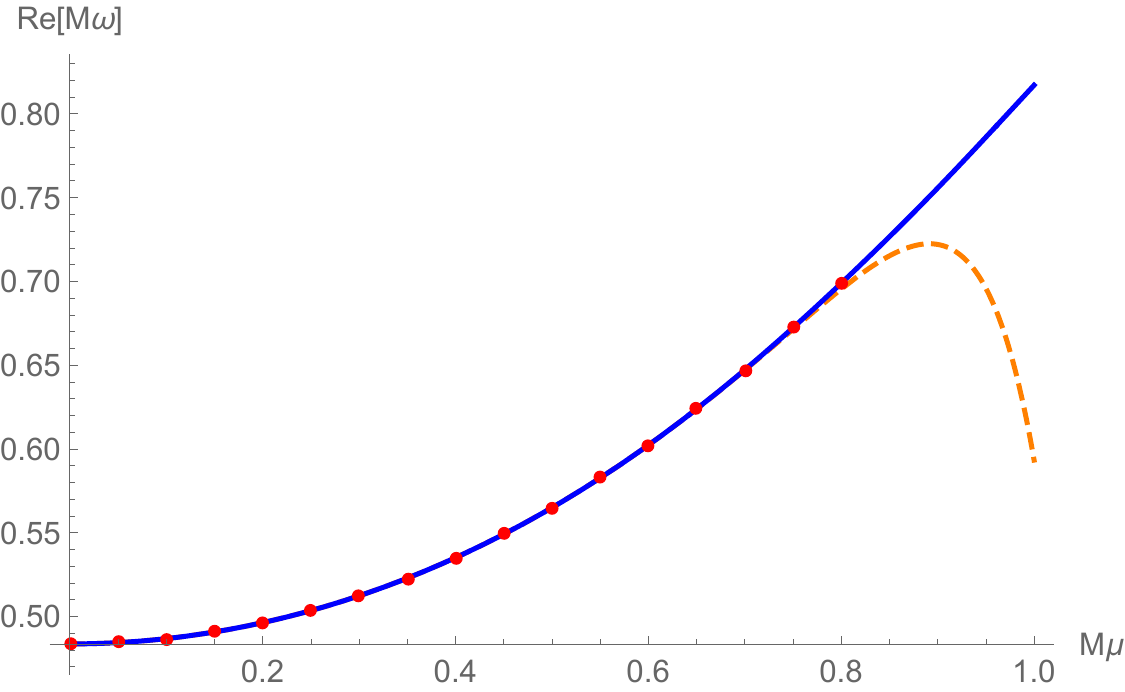}
  \end{minipage} \hspace{1cm}
  \begin{minipage}[b]{0.45\linewidth}
    \centering
    \includegraphics[width=0.95\linewidth]{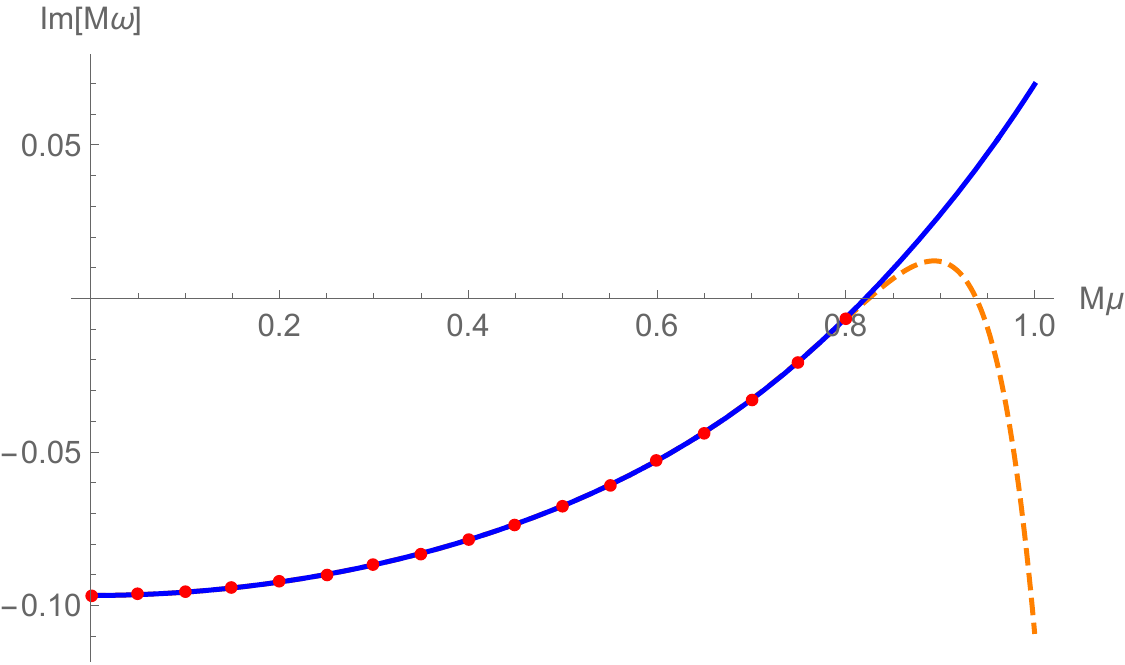}
  \end{minipage} 
\end{center}
  \caption{The mass dependence of the $\ell=2$ fundamental QNM frequency for the massive scalar perturbation. The (red) points represent the numerical values. The (orange) dashed line and the (blue) solid line are the perturbative series in Eq.~\eqref{eq:massive-pert} up to $k=40$ and its diagonal Pad\'e approximant, respectively. The Pad\'e approximant is extrapolated beyond the radius of convergence.}
  \label{fig:massiveQNM}
\end{figure}

By using the Pad\'e approximants, we finally extrapolate our perturbative results to the finite parameter region, as shown in Figure~\ref{fig:massiveQNM}.

\subsection{Slowly rotating black holes}
Another simple application is the Kerr geometry. We regard the angular momentum as a deformation parameter.
Here we consider the slow rotation limit. We briefly explain how to get the slow rotation expansion of the QNM frequency reported in \cite{hatsuda2021}.

The perturbation of the rotating black holes are governed by the Teukolsky equation \cite{teukolsky1972}.
In \cite{hatsuda2021}, an isospectral equation to the Teukolsky equation was proposed.
This isospectral equation is much more useful for our purpose in this paper.
We start with the radial master equation
\begin{equation}
\begin{aligned}
\left( \frac{d^2}{dx^2}+(2M\omega)^2-V(x) \right) \Phi(x)=0,
\end{aligned}
\label{eq:alternative-Kerr}
\end{equation}
where
\begin{equation}
\begin{aligned}
V(x)&=f(z)\biggl[ 4c^2+\frac{4c(m-c)}{z}+\frac{{}_sA_{\ell m}(c)+s(s+1)-c(2m-c)}{z^2}-\frac{s^2-1}{z^3}\biggr],\\
f(z)&=1-\frac{1}{z},\qquad
x=z+\log(z-1)
\end{aligned}
\label{eq:alternative-Kerr-2}
\end{equation}
and $c=a\omega$ is related to the rotation parameter $a$.
For the notational detail, see \cite{hatsuda2021}.
Of course, the slow rotation limit corresponds to the small $c$ limit.
The separation constant ${}_sA_{\ell m}(c)$ is determined by the regularity condition of the angular master equation at $\xi=\pm 1$:
\begin{equation}
\begin{aligned}
\biggl[ \frac{d}{d\xi}(1-\xi^2)\frac{d}{d\xi}+(c \xi)^2-2cs \xi+{}_sA_{\ell m}(c)+s-\frac{(m+s\xi)^2}{1-\xi^2} \biggr] {}_sS_{\ell m}(\xi)=0.
\end{aligned}
\label{eq:angular-Kerr}
\end{equation}
To compute the small $c$ expansion of the potential, we need the perturbative series of ${}_sA_{\ell m}(c)$. This can be done as follows.
In $c \to 0$, the angular master equation can be solved exactly. The regular solution at $\xi=\pm 1$ exists only for the discrete eigenvalue
\begin{equation}
\begin{aligned}
{}_sA_{\ell m}(0)&=\ell(\ell+1)-s(s+1),
\end{aligned}
\end{equation}
and the exact eigenfunction is given by
\begin{equation}
\begin{aligned}
{}_sS_{\ell m}^{(c=0)}(\xi)&=
(1-\xi)^{-\frac{m+s}{2}}(1+\xi)^{\frac{m-s}{2}} P_{\ell+s}^{(-m-s,m-s)}(\xi),
\end{aligned}
\end{equation}
where $P_n^{(\alpha,\beta)}(z)$ is the Jacobi polynomial. We have assumed $\ell \geq |s|$ and $|m| \leq \ell$. 
As in the very similar treatment in the Bender-Wu approach, the eigenvalue ${}_sA_{\ell m}(c)$ and the eigenfunction ${}_sS_{\ell m}(\xi)$ admit the perturbative series in $c$:
\begin{equation}
\begin{aligned}
{}_sA_{\ell m}(c)=\sum_{k=0}^\infty c^k {}_sA_{\ell m}^{(k)},\qquad
{}_sS_{\ell m}(\xi)=\sum_{k=0}^\infty c^k {}_sS_{\ell m}^{(k)}(\xi).
\end{aligned}
\end{equation}
The crucial step is to find the following general structure of the regular function ${}_sS_{\ell m}^{(k)}(\xi)$:
\begin{equation}
\begin{aligned}
{}_sS_{\ell m}^{(k)}(\xi)=(1-\xi)^{-\frac{m+s}{2}}(1+\xi)^{\frac{m-s}{2}} {}_sQ_{\ell m}^{(k)}(\xi),
\end{aligned}
\end{equation}
where ${}_sQ_{\ell m}^{(k)}(\xi)$ is a polynomial of degree $\ell+s+k$ in $\xi$. From the differential equation \eqref{eq:angular-Kerr}, we can fix all the coefficients in the polynomial ${}_sQ_{\ell m}^{(k)}(\xi)$ and ${}_sA_{\ell m}^{(k)}$ order by order.
This method allows us to compute the exact value of ${}_sA_{\ell m}^{(k)}$ up to very high orders for given $s$, $\ell$ and $m$. We have confirmed that the first few coefficients indeed agree with the results in \cite{berti2006, berti2006a}.

Once we know the small $c$ expansion of ${}_sA_{\ell m}(c)$, we obtain the perturbative expansion of the potential $V(x)$. Then we can apply the method in Section~\ref{sec:general}. The result is given by the following  small $c$ expansion:
\begin{equation}
\begin{aligned}
M {}_s\omega_{\ell m}=\sum_{k=0}^\infty c^k {}_sv_{\ell m}^{(k)}.
\end{aligned}
\label{eq:omega-c}
\end{equation}
However, we are interested in the perturbative expansion in terms of the rotation parameter $a$ rather than $c=a\omega$.
This expansion is easily obtained by plugging Eq.~\eqref{eq:omega-c} into $c=a\omega$ and by inversely expanding $c$ in $a/M$.
We finally obtain the following perturbative series
\begin{equation}
\begin{aligned}
M {}_s\omega_{\ell m}=\sum_{k=0}^\infty \left( \frac{a}{M} \right)^k {}_sw_{\ell m}^{(k)},
\end{aligned}
\label{eq:omega-a}
\end{equation}
where the explicit values of ${}_sw_{\ell m}^{(k)}$ for $(s, \ell, m)=(-2, 2, 0), (-2, 2, 1), (-2, 2, 2)$ up to $k=12$ are found in Table~1 in \cite{hatsuda2021}.

\subsection{Almost asymptotically flat black holes}
We can also apply our formalism to asymptitically non-flat geometries. 
We focus on the Schwarzschild de Sitter black holes.
In this case, the functions in the minimally coupled massless scalar/vector/odd-parity gravitational perturbations are all given by
\begin{equation}
\begin{aligned}
f(r)&=1-\frac{2M}{r}-\frac{\Lambda r^2}{3}, \\
V(x)&=f(r)\( \frac{\l(\l+1)}{r^2}+(1-s^2)\(\frac{2M}{r^3}-\frac{4-s^2}{6}\Lambda \) \),
\end{aligned}
\label{eq:dS-funcs}
\end{equation}
where $s=0,1,2$ denotes the spin-weight of the perturbation fields, and $\Lambda$ is the cosmological costant.
We regard $\Lambda$ as a deformation parameter.
In contrast to the previous examples, the function $f(r)$ depends on $\Lambda$.
The explicit relation between $r$ and $x$ is now quite complicated.
As discussed in Section \ref{sec:general}, we have to use the relation in Eq.~\eqref{eq:tortoise-3} to eliminate $r$.
This can be done at least perturbatively with respect to $\Lambda$.
After this prescription, the potential in terms of $x$ receives an infinite number of perturbative corrections.
We apply the Bender-Wu approach for such a perturbative series of the potential.
In the Bender-Wu approach, we need the Taylor series of the perturbative corrections to the potential around the extremal point $x=\bar{x}$ of the zeroth potential.
This can be done systematically.

We expand the frequency as
\begin{equation}
\begin{aligned}
M\omega=\sum_{k=0}^\infty (9M^2 \Lambda)^{k} w_k.
\end{aligned}
\label{eq:dS}
\end{equation}
The numerical values of $w_k$ for the fundamental mode with $\l=2$ in the gravitational perturbation ($s=2$) up to $k=8$ are given in Table~\ref{tab:dS}.

\begin{table}[tp]
\caption{The first eight perturbative corrections to the fundamental QNM frequency in Eq.~\eqref{eq:dS} with $\l=2$ in the odd-parity gravitational perturbation for the asymptotically dS black holes. It turns out that the same values are also obtained by the even-parity perturbation.}
\begin{center}
\begin{tabular}{c@{\hspace{10pt}}c}
\hline
$k$ & $w_k$ \\
\hline
$0$ &  $0.3736716844 - 0.0889623157i$       \vspace{-5pt}\\
$1$ &  $-0.1864855559 + 0.0372042528i$     \vspace{-5pt}\\
$2$ &  $-0.04819480629 + 0.01428258071i$   \vspace{-5pt}\\
$3$ &  $-0.02302643485 + 0.00713463072i$    \vspace{-5pt}\\
$4$ &  $-0.01415049627 + 0.00398414719i$    \vspace{-5pt}\\
$5$ &  $-0.010032759238 + 0.002550521089i$       \vspace{-5pt}\\
$6$ &  $-0.007668666891 + 0.001893042626i$     \vspace{-5pt}\\
$7$ & $-0.006085692144 + 0.001548612387i$ \vspace{-5pt}\\
$8$ & $-0.004939500648 + 0.001314426006i$ \\
\hline
\end{tabular}
\end{center}
\label{tab:dS}
\end{table}%

A non-trivial test of our result is to check the isospectrality between the odd-parity and even-parity gravitational perturbations.
The potential in the even-parity gravitational perturbation is
\begin{equation}
\begin{aligned}
V^\text{even}(x)=f(r)\frac{2}{r^3} \frac{9M^3+3\lambda^2 M r^2+\lambda^2(1+\lambda) r^3+9M^2 \lambda r-3M^2 \Lambda r^3}{(3M+\lambda r)^2},
\end{aligned}
\end{equation}
where $\lambda=(\l-1)(\l+2)/2$.
It is well-known that the QNM spectra in the odd/even-parity perturbations are exactly same.
The reason behind this remarkable fact is a supersymmetric structure. See appendix A in \cite{berti2009}.
Our formalism is also applicable to this potential, and
we have checked that the isospectrality indeed holds at the perturbative level at least up to $k=8$:
\begin{equation}
\begin{aligned}
w_k^\text{odd}=w_k^\text{even}.
\end{aligned}
\end{equation}
This is an evidence of the validity of our method.

Let us discuss the extrapolation of Eq.~\eqref{eq:dS} to finite $\Lambda$.
We first observe that the perturbative series is likely convergent, but it is hard to guess the radius of convergence from the coefficient $w_k$. 
We consider the $[4/4]$ Pad\'e approximant by using the values in Table~\ref{tab:dS}.
The Pad\'e approximant $\omega^{[4/4]}$ for $(s,\l)=(2,2)$ has four poles at
\begin{equation}
\begin{aligned}
M^2\Lambda=0.101-0.0134i, \quad 0.142+0.00389i,\quad 0.323+0.0678i,\quad 2.45+0.687i,
\end{aligned}
\end{equation}
where the first pole is relatively close to $M^2\Lambda=1/9$, at which the event horizon and the de Sitter horizon coincide. It is expected that higher-order Pad\'e approximants capture this observation more precisely, but it is technically difficult to check it at the moment. This observation implies that the radius of convergence of Eq.~\eqref{eq:dS} is just $|M^2 \Lambda|=1/9$.

The extrapolation of Eq.~\eqref{eq:dS} by its Pad\'e approximant is compared to the numerical value of the QNM frequency directly computed from Eq.~\eqref{eq:dS-funcs}.
For $M^2\Lambda=0.06$, we have
\begin{equation}
\begin{aligned}
M\omega^{[4/4]}_{s=2,\l=2}(M^2\Lambda=0.06) \approx 0.2533-0.06304i,
\end{aligned}
\end{equation}
which agrees with the WKB result in \cite{zhidenko2003} and also a recent high-precision computation in \cite{hatsuda2020d}.

\begin{figure}[tb]
\begin{center}
  \begin{minipage}[b]{0.45\linewidth}
    \centering
    \includegraphics[width=0.95\linewidth]{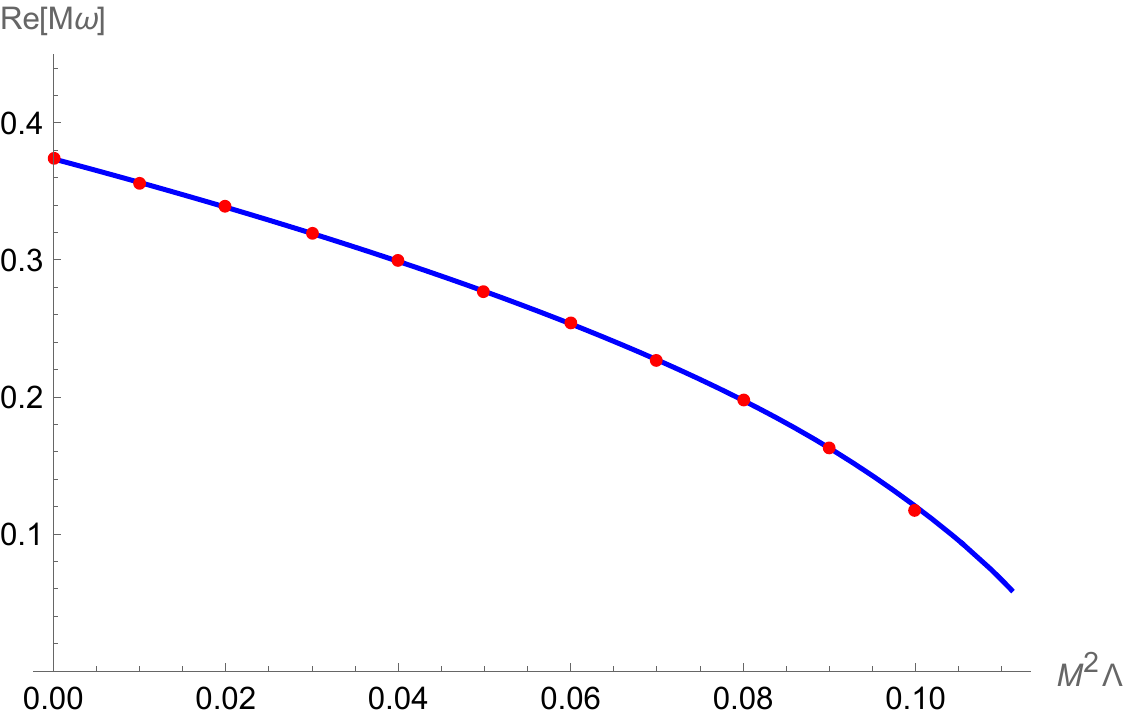}
  \end{minipage} \hspace{1cm}
  \begin{minipage}[b]{0.45\linewidth}
    \centering
    \includegraphics[width=0.95\linewidth]{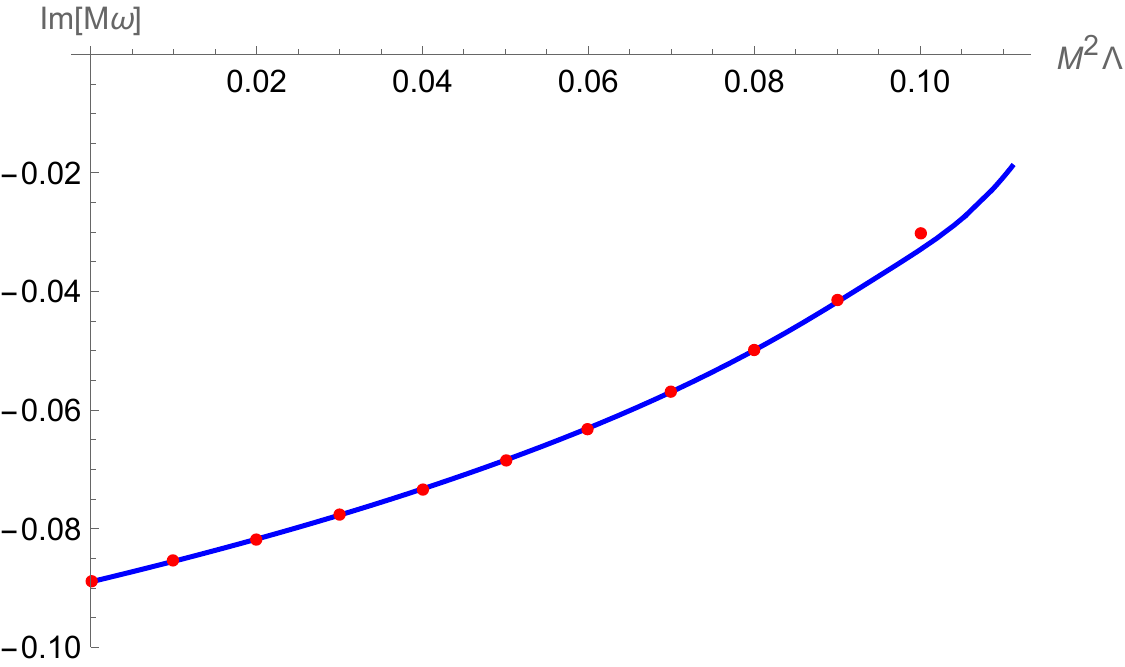}
  \end{minipage} 
\end{center}
  \caption{The cosmological constant dependence of the fundamental QNM frequency for the asymptotically Schwarzschild de Sitter black holes. The (red) points represent the numerical values, while the (blue) solid line represents the $[4/4]$ Pad\'e approximant.}
  \label{fig:dS}
\end{figure}

We should note that the QNM spectral problem becomes quite different for $\Lambda>0$ (dS) and $\Lambda<0$ (AdS).
The boundary condition in the AdS case is much more involved than the dS case \cite{cardoso2001, berti2009}.
In this paper, we restrict ourselves to the dS case for simplicity. It would be interesting to clarify a physical meaning of a na\"ive continuation to $\Lambda<0$ of our result.
Another perturbative treatment for the (A)dS spectral problem will be also found in \cite{Aminov:2023jve}.

\subsection{Reissner-Nordstr\"om black holes}
The spectrum for the Reissner-Nordstr\"om black holes are more involved.
The master equation in the odd-parity gravitational perturbation consists of
\begin{equation}
\begin{aligned}
f(r)&=1-\frac{2M}{r}+\frac{Q^2}{r^2},\\
V(x)&=f(r)\( \frac{\l(\l+1)}{r^2}-\frac{q}{r^3}+\frac{4Q^2}{r^4} \),
\end{aligned}
\end{equation}
where
\begin{equation}
\begin{aligned}
q=3M+\sqrt{9M^2+4Q^2(\l-1)(\l+2)}.
\end{aligned}
\end{equation}
We have two characteristic regimes: $Q=0$ and $Q=M$.
We discuss perturbative series around these two points.

\subsubsection{Almost chargeless limit}
The first is the small charge expansion.
In this case, $Q^2$ is a natural deformation parameter.
We write the perturbative QNM frequency as
\begin{equation}
\begin{aligned}
M \omega=\sum_{k=0}^{\infty} \( \frac{Q}{M} \)^{2k} w_k.
\end{aligned}
\label{eq:RN-pert}
\end{equation}
The potential receives an infinite number of corrections.
The strategy is the same as that in the previous subsection.
We show the numerical values of the perturbative coefficients $w_k$ for the fundamental QNM frequency with $\l=2$ up to $k=4$ in Table~\ref{tab:RN}.
The quadratic correction $w_1$ matches well with \cite{cardoso2019}.

\begin{table}[tp]
\caption{The low order corrections to the fundamental QNM frequency for $\l=2$ in the Reissner-Nordstr\"om gravitational perturbation. We consider the two distinct perturbative series \eqref{eq:RN-pert} and \eqref{eq:RN-ext}.}
\begin{center}
\begin{tabular}{c@{\hspace{10pt}}c@{\hspace{10pt}}c}
\hline
$k$ & $w_k$ & $w_k^\text{ext}$ \\
\hline
$0$ &  $0.3736716844 - 0.0889623157i$    & $0.4313408007-0.0834603151 i$   \vspace{-5pt}\\
$1$ &  $0.02581767285 - 0.00282403214i$   & $-0.2070138464-0.0853606869 i$  \vspace{-5pt}\\
$2$ &  $0.02518778870 + 0.00020532453i$  & $0.2543444995+0.4939946909 i$ \vspace{-5pt}\\
$3$ &  $-0.004748170246 + 0.002508402108 i$ & $0.758606111-1.429576400 i$   \vspace{-5pt}\\
$4$ &  $0.01557265014 + 0.00041287974i$   & $-6.158687644 + 0.575432188 i$  \\
\hline
\end{tabular}
\end{center}
\label{tab:RN}
\end{table}%

\subsubsection{Almost extremal limit}
We can also consider another limit $Q \to M$.
In this case, $1-Q/M$ is a good parameter.
Therefore we write the frequency as
\begin{equation}
\begin{aligned}
M\omega=\sum_{k=0}^\infty \alpha^k w_k^\text{ext},\qquad \alpha:=1-\frac{Q}{M}.
\end{aligned}
\label{eq:RN-ext}
\end{equation}
Now we have
\begin{equation}
\begin{aligned}
f(r)=\left(1-\frac{M}{r}\right)^2-\alpha \frac{2M^2}{r^2}+\alpha^2 \frac{M^2}{r^2}.
\end{aligned}
\end{equation}
We also expand the potential perturbatively with respect to $\alpha$.
The QNM frequencies in the strictly extremal case ($\alpha=0$) can be computed by the Bender-Wu approach \cite{hatsuda2020}.
We do the same computation for high-order corrections.
The numerical values of $w_k^\text{ext}$ for the fundamental QNM frequency with $\l=2$ up to $k=4$ are shown in Table~\ref{tab:RN}.
The zeroth order coefficients $w_0^\text{ext}$ agrees with the early result \cite{onozawa1996}. We do not find any references on the perturbative corrections near the extremal limit. 

\subsubsection{An interpolating function}
\begin{figure}[tb]
\begin{center}
  \begin{minipage}[b]{0.45\linewidth}
    \centering
    \includegraphics[width=0.95\linewidth]{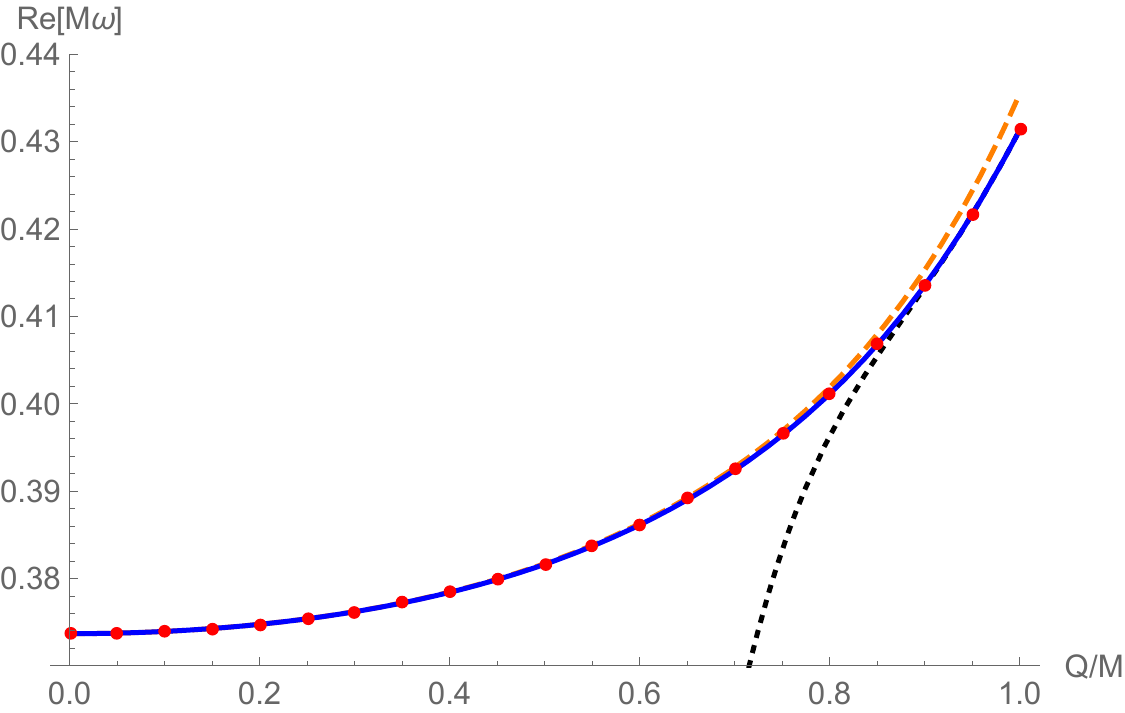}
  \end{minipage} \hspace{1cm}
  \begin{minipage}[b]{0.45\linewidth}
    \centering
    \includegraphics[width=0.95\linewidth]{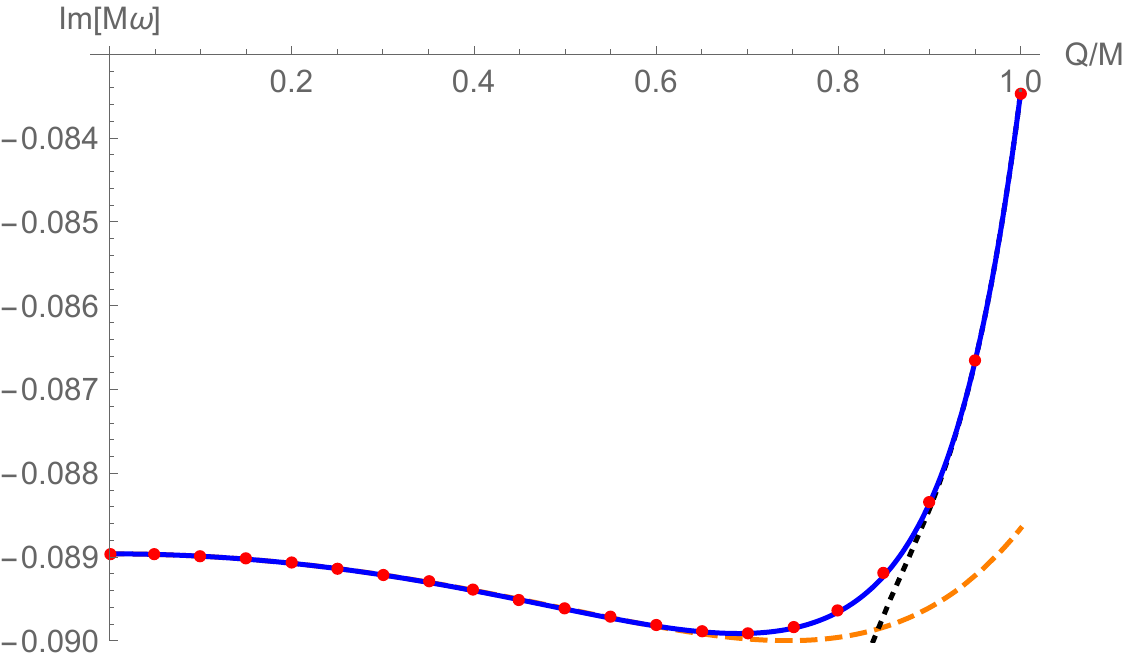}
  \end{minipage} 
\end{center}
  \caption{The (red) points represent the numerical values of the QNM frequency of the RN black holes. The (blue) solid curve is the graph of the rational function \eqref{eq:interpolation} for $p=q=4$ with the coefficients in Table~\ref{tab:rational}. The (orange) dashed and (black) dotted lines represent the perturbative expansions \eqref{eq:RN-pert} and \eqref{eq:RN-ext} up to $k=4$, respectively.}
  \label{fig:interpolation}
\end{figure}

We have the two perturbative expansions of the same spectrum in the different regimes.
In each regime, we determine its Pad\'e approximant, and can extrapolate it to the other regime.
However, to know the global behavior, there is a better approximation, called multi-point Pad\'e approximants \cite{bender1978, bakerjr.1996}.
Let us consider a rational function
\begin{equation}
\begin{aligned}
M\omega^{[p/q]}=\frac{a_0+a_1 Q/M+\cdots +a_p (Q/M)^p}{1+b_1 Q/M+\cdots +b_q (Q/M)^q}.
\end{aligned}
\label{eq:interpolation}
\end{equation}
We fix the coefficients $a_n$ and $b_n$ so that the rational function reproduces the \textit{both} perturbative expansions around $Q/M=0$ and $Q/M=1$. 
For instance to get the rational function $M\omega^{[4/4]}$ we totally need nine data in Eq.~\eqref{eq:RN-pert} and in Eq.~\eqref{eq:RN-ext}.
A balanced choice is to take $w_k$ ($0 \leq k \leq 2$) in Eq.~\eqref{eq:RN-pert} and $w_k^\text{ext}$ ($0 \leq k \leq 3$) in Eq.~\eqref{eq:RN-ext}.
Recall the expansion Eq.~\eqref{eq:RN-pert} has no odd-order terms. We can use this information to fix $a_n$ and $b_n$.
The explicit values of $a_n$ and $b_n$ in this case are shown in Table~\ref{tab:rational}.
The interpolating function remarkably reproduces the numerical values in the whole regime $0\leq Q/M \leq 1$, as shown in Figure~\ref{fig:interpolation}.

Interpolating functions will be improved if one considers further perturbative expansions around other points in the middle region.
For instance, a perturbative expansion around $Q/M \sim 0.8$ will provide us an important information on the global structure of the imaginary part of the QNM frequency for $\l=2$.
We do not compute it in this work, but expect that our method is still applicable in such situations.

\begin{table}[tp]
\caption{The nine coefficients in the rational approximation $M\omega^{[4/4]}$ for the $\l=2$ fundamental mode.}
\begin{center}
\begin{tabular}{c@{\hspace{10pt}}c@{\hspace{10pt}}c}
\hline
$n$  &   $a_n$    &     $b_n$   \\ 
\hline
$0$  & $0.3736716844 - 0.0889623157i$  &   \vspace{-5pt}\\
$1$  & $-0.349769907 + 0.062882011i$   &  $-0.92374126 - 0.051639318i$ \vspace{-5pt}\\
$2$  & $-0.342112665 - 0.038824176i$   &  $-0.91011322 - 0.313017895i$ \vspace{-5pt}\\
$3$  & $0.492170748 - 0.023942699i$   &  $1.32244473 + 0.24735505i$ \vspace{-5pt}\\
$4$  & $-0.169504965 + 0.033347865i$   &  $-0.454637541 - 0.004795289i$ \\ 
\hline  
\end{tabular}
\end{center}
\label{tab:rational}
\end{table}%

\subsection{Parameterized black hole QNMs}\label{subsec:para}
Recently, a simple and effective way to compute perturbative corrections was proposed in \cite{cardoso2019, mcmanus2019, kimura2020}.
We refer to it as the \textit{parameterized QNM approach}.
As one can see in the previous examples, most deformation terms in the potential take the form as linear combinations of $1/r^j$ with integral $j$.
At the first order in the perturbation, corrections to the QNM frequencies are the same linear combinations of the potential. See Eqs.~\eqref{eq:multi-para-pot}
and \eqref{eq:multi-para}. The main idea of the parameterized QNM approach is the following. We make a list of corrections generated by only the $1/r^j$-deformations beforehand, and use it for a more complicated potential to which corrections are linear combinations of the $1/r^j$-deformations. The extension to high-order corrections is straightforward \cite{mcmanus2019}.
Physical applications of the parameterized QNM approach have been shown in~\cite{Tattersall:2019nmh, hatsuda2020b, deRham:2020ejn, Volkel:2022aca, Volkel:2022khh, Franchini:2022axs, Lahoz:2023csk, Ghosh:2023etd, Mukohyama:2023xyf, Franchini:2023eda}.
(See also Appendix.~\ref{app:ejidentities} for complementary discussion.)

At the technical level, it is not so easy to compute the precise values of the quadratic corrections.
In \cite{cardoso2019, mcmanus2019}, the authors used numerical fittings.
Since our formalism is easily applied to the setup of the parameterized QNM approach, we re-evaluate the corrections up to the quadratic order.
This re-evaluation played an important role in the computation of perturbative corrections for slowly rotating black holes \cite{hatsuda2020b}.
We keep at least ten-digit precision for all the corrections listed in this section.
We focus on deformation of the odd-parity gravitational perturbation of the Schwarzschild black holes.
The computations for the other cases are straightforward.
The potential is
\begin{equation}
\begin{aligned}
V_0(x)=f(r)\( \frac{\l(\l+1)}{r^2}-\frac{3r_H}{r^3} \),\quad
V_1(x)=\frac{f(r)}{r_H^2} \left(\frac{r_H}{r}\right)^j,\quad V_{k\geq 2}(x)=0,
\end{aligned}
\end{equation}
where $r_H$ is the location of the event horizon and $j=0,1,2,\dots$.
For this deformation, the spectrum receives the corrections:
\begin{equation}
\begin{aligned}
\omega=\omega_0+\sum_{k=1}^\infty \alpha^k e_j^{(k)}.
\end{aligned}
\end{equation}
For $\ell=2$, we show the numerical values of $e_j^{(k)}$ ($0\leq j \leq 8$, $k=1,2$) in Table~\ref{tab:diagonal}.

\begin{table}[tp]
\caption{The one-parameter corrections up to the second order for $\l=2$ in the parameterized QNM approach.}
\begin{center}
\begin{tabular}{c@{\hspace{10pt}}c@{\hspace{10pt}}c}
\hline
$k$ & $j$ &  $r_H e_j^{(k)}$ \\
\hline
	& $0$ &  $0.2472519654+0.0926430738i$       \vspace{-5pt}\\
       & $1$ &  $0.1598547870 + 0.0182084818i$     \vspace{-5pt}\\
       & $2$ &  $0.09663224013 - 0.00241549645i$   \vspace{-5pt}\\
       & $3$ &  $0.05849078501 - 0.00371786129i$    \vspace{-5pt}\\
$1$ & $4$ &  $0.03667943678 - 0.00043869695i$    \vspace{-5pt}\\
       & $5$ & $0.02403794775 + 0.00273079314i$  \vspace{-5pt}\\
       & $6$ & $0.01634281096 + 0.00484267168i$   \vspace{-5pt}\\
       & $7$ & $0.011363575081 + 0.006013991932i$  \vspace{-5pt}\\
       & $8$ & $0.007951997735 + 0.006536996457i$  \\
\hline
	& $0$ &  $0.002868401222 - 0.001011345890i$  \vspace{-5pt}\\
       & $1$ &  $-0.01439027937 - 0.00572350838 i$       \vspace{-5pt}\\
       & $2$ &  $-0.005756554781 + 0.000336740545i$ \vspace{-5pt}\\
       & $3$ &  $-0.0006273259154 - 0.0004693348600i$ \vspace{-5pt}\\
$2$ & $4$ &  $0.0007234494450 - 0.0011595941966i$  \vspace{-5pt}\\
       & $5$ & $0.000987182421 - 0.001122519006i$  \vspace{-5pt}\\
       & $6$ & $0.0010046849768 - 0.0008403243677i$  \vspace{-5pt}\\
       & $7$ & $0.0009526541187 - 0.0005456646402i$ \vspace{-5pt}\\
       & $8$ & $0.0008715569057 - 0.0003017937415i$\\
\hline
\end{tabular}
\end{center}
\label{tab:diagonal}
\end{table}%

To make a list at the quadratic order, we also have to consider two-parameter perturbations in Eq.~\eqref{eq:multi-para-pot} with
\begin{equation}
\begin{aligned}
V_1^{\alpha}(x)&=\frac{f(r)}{r_H^2}\left( \frac{r_H}{r} \right)^i,\qquad V_{k\geq 2}^{\alpha}(r)=0, \\
V_1^{\beta}(x)&=\frac{f(r)}{r_H^2}\left( \frac{r_H}{r} \right)^j,\qquad V_{k\geq 2}^{\beta}(r)=0.
\end{aligned}
\end{equation}
For this perturbation, the frequency receives the corrections:
\begin{equation}
\begin{aligned}
\omega=\omega_0+\sum_{k=1}^\infty \sum_{\ell=0}^k \alpha^\ell \beta^{k-\ell} e_{ij}^{(\ell, k-\ell)}.
\end{aligned}
\end{equation}
where we have $e_{ij}^{(k,0)}=e_i^{(k)}$ and $e_{ij}^{(0,k)}=e_j^{(k)}$ by construction.
Therefore at the second order, the only unknown coefficient is $e_{ij}^{(1,1)}$.
This can be evaluated by the trick explained in Eq.~\eqref{eq:multi-para}.
The numerical values are shown in Table~\ref{tab:off-diagonal}.
We compare these results with \cite{cardoso2019, mcmanus2019}, and found that there are significant differences.

For the error estimation of the coefficients in Tables~\ref{tab:diagonal} and~\ref{tab:off-diagonal}, 
we use the recursion relations among the coefficients in Eqs.~\eqref{eq:recrelation1st}
and~\eqref{eq:recrelation2nd}.
We checked that 
Eq.~\eqref{eq:recrelation1st} is satisfied at ${\cal O}(10^{-15})$ for linear coefficients, 
and Eq.~\eqref{eq:recrelation2nd} is satisfied at ${\cal O}(10^{-11})$ for quadratic coefficients,
while those equations are satisfied at ${\cal O}(10^{-6})$ and ${\cal O} (10^{-2})$, respectively, in the previous works~\cite{cardoso2019, mcmanus2019}.
This also shows that the perturbative approach developed in the present paper works well.

\begin{table}[tp]
\caption{The off-diagonal quadratic corrections for $\l=2$ in the two-parameter perturbation.}
\begin{center}
\begin{tabular}{c@{\hspace{10pt}}c@{\hspace{10pt}}c}
\hline
$i$ & $j$ &  $r_H e_{ij}^{(1,1)}$ \\
\hline
$0$ & $1$ & $-0.02588238896 - 0.02792966573i$   \vspace{-5pt}\\
       & $2$ & $-0.03870432587 - 0.02320896618i$    \vspace{-5pt}\\
       & $3$ & $-0.03739171923 - 0.01523959074i$    \vspace{-5pt}\\
       & $4$ & $-0.03119143980 - 0.01062473399 i$    \vspace{-5pt}\\
       & $5$ & $-0.02473633363 - 0.00886735210i$    \vspace{-5pt}\\
       & $6$ & $-0.01939362275 - 0.00853499059i$    \vspace{-5pt}\\
       & $7$ & $-0.01523819641 - 0.00870770164i$   \\
\hline
$1$ & $2$ &     $-0.02293084111 - 0.00341311941i$  \vspace{-5pt}\\
       & $3$ &  $-0.01688392216 - 0.00102025764i$   \vspace{-5pt}\\
       & $4$ &  $-0.01249473743 - 0.0009507495878i   $ \vspace{-5pt}\\
       & $5$ &  $-0.009533459569 - 0.001537281111i$    \vspace{-5pt}\\
       & $6$ & $-0.007497937650 - 0.002167588509i$    \vspace{-5pt}\\
       & $7$ & $-0.006026376454 - 0.002674270477i$    \\
\hline
$2$ & $3$ &     $-0.005785247726+ 0.0002460429730i    $ \vspace{-5pt}\\
	& $4$ & $-0.003236295992- 0.0006934041512i   $  \vspace{-5pt}\\
       & $5$ &  $-0.002075023229 - 0.001296233004i$    \vspace{-5pt}\\
       & $6$ & $-0.001466727846 - 0.001581112859i$    \vspace{-5pt}\\
       & $7$ & $-0.001083345654 - 0.001682173863i$    \\
\hline
$3$ & $4$ &  $0.000315183631 - 0.001771852361i$  \vspace{-5pt}\\       
       & $5$ & $0.000806605055 - 0.002028059473i$    \vspace{-5pt}\\
       & $6$ & $0.000954987015 - 0.001956886197i$    \vspace{-5pt}\\
       & $7$ & $0.001002044048 - 0.001756226616i$    \\
\hline
$4$ & $5$ &  $0.001737194187 - 0.002338806036i$  \vspace{-5pt}\\
    & $6$ & $0.001773835947 - 0.002098671851i$    \vspace{-5pt}\\
    & $7$ & $0.001741580709 - 0.001777161054i$    \\
\hline
$5$ & $6$ &  $0.001993021672 - 0.001958873499i$  \vspace{-5pt}\\
    & $7$ & $0.001947091748 - 0.001620453116i$    \\
\hline
$6$ & $7$ &  $0.001959730533 - 0.001367519282i$  \\
\hline
\end{tabular}
\end{center}
\label{tab:off-diagonal}
\end{table}%

\subsection{Series expansion method}\label{app:series}
As an application of our perturbative framework based on a method other than the Bender-Wu approach,
we study the series expansion method known as Leaver's method~\cite{leaver1985, konoplya2011}.
We consider the system with the parameterized QNM potential in Eqs~\eqref{eq:parameterizedpotential}-\eqref{eq:parameterizedA} 
with a single correction term
\begin{align}
\delta V = \alpha \frac{f}{r_H^2}\left(\frac{r_H}{r}\right)^j,
\end{align}
where $f$ is given by $f = 1 - r_H/r$.
We assume the following series expansion of the wave function as
\begin{align}
\Phi = e^{i \omega r_* } \sum_{k = 0}^\infty a_k f^{k + n},
\end{align}
where the characteristic exponent $n$ is given by 
\begin{align}
n = - 2 i r_H \omega,
\end{align}
so that the QNM boundary condition at $r = r_H$ is satisfied.
After some calculations, we obtain recursion relations for $a_k$
\begin{align}
A_k a_{k-1} + B_k a_k + C_k a_{k+1} 
+ \alpha  \sum_{m = 0}^{j-2}  
D_m 
a_{k-m} 
=0,
\label{recrelappendixb}
\end{align}
where coefficients $A_k, B_k, C_k$ and $D_m$ are given by
\begin{align}
A_k &= (k-2 - 2 i r_H \omega)(k+2 - 2 i r_H \omega),
\\
B_k &= 3 - 2 k (1+k) - \ell(\ell + 1) 
+ 4 i r_H \omega( 1 + 2  k )
+ 8 r_H^2 \omega^2,
\\
C_k &= (1 + k)(1 + k - 2 i r_H \omega),
\\
D_m &= \frac{(-1)^{m+1} (j-2)!}{m!(j-2-m)!}.
\end{align}
The coefficients $a_k$ with large $k$ take exponentially small value 
only for the wave function with the appropriate QNM boundary condition at $r \to \infty$.
Thus, we can calculate the approximate QNM frequency by setting 
\begin{align}
a_{k_{\rm max}} = 0,
\label{eq:akmax}
\end{align}
with a large integer $k_{\rm max}$.
However, directly solving Eq.~\eqref{eq:akmax} numerically is
very difficult, and then we usually use 
Leaver's continued fraction method~\cite{leaver1985, konoplya2011} whose basic equation 
is mathematically same as Eq.~\eqref{eq:akmax}.
In this section, we study this problem based on our perturbative approach.

Expanding the coefficients $a_k$ and the QNM frequency $\omega$ as
\begin{align}
a_k &= a_k^{(0)} + \alpha a_k^{(1)} + \alpha^2 a_k^{(2)} + \cdots,
\\
\omega &= \omega_0 + \alpha \omega_1 + \alpha^2 \omega_2 + \cdots,
\end{align}
the coefficients $A_k, B_k, C_k$ become
\begin{align}
A_k &= A_k^{(0)} + \alpha \omega_1 A_k^{(1)} + \alpha^2 \omega_1^2 A_k^{(2,0)} + \alpha^2 \omega_2 A_k^{(0,1)}  + \cdots,
\\
B_k &= B_k^{(0)} + \alpha \omega_1 B_k^{(1)} + \alpha^2 \omega_1^2 B_k^{(2,0)} + \alpha^2 \omega_2 B_k^{(0,1)} + \cdots,
\\
C_k &= A_k^{(0)} + \alpha \omega_1 C_k^{(1)}+ \alpha^2 \omega_1^2 C_k^{(2,0)} + \alpha^2 \omega_2 C_k^{(0,1)} + \cdots,
\end{align}
where the coefficients in RHS depend only on $\omega_0$.
The recursion relations in Eq.~\eqref{recrelappendixb} at each order become 
\begin{align}
{\cal O}(\alpha^0): \quad & A_k^{(0)} a_{k-1}^{(0)} + B_k^{(0)} a_k^{(0)} + C_k^{(0)} a_{k+1}^{(0)} = 0,
\label{eq:recrelappendixb0th}
\\
{\cal O}(\alpha^1): \quad & 
A_k^{(0)} a_{k-1}^{(1)} + B_k^{(0)} a_k^{(1)} + C_k^{(0)} a_{k+1}^{(1)}
\notag\\ & +
\omega_1 
\Big[
A_k^{(1)} a_{k-1}^{(0)} + B_k^{(1)} a_k^{(0)} +C_k^{(1)} a_{k+1}^{(0)} 
\Big]
+ 
\sum_{m = 0}^{j-2} D_m
a_{k-m}^{(0)} 
=0,
\label{eq:recrelappendixb1st}
\\
{\cal O}(\alpha^2): \quad &  
A_k^{(0)} a_{k-1}^{(2)} + B_k^{(0)} a_k^{(2)} + C_k^{(0)} a_{k+1}^{(2)}
\notag\\& +
\omega_1 
\Big[
A_k^{(1)} a_{k-1}^{(1)} + B_k^{(1)} a_k^{(1)} +C_k^{(1)} a_{k+1}^{(1)} 
\Big]
\notag\\& +
\omega_1^2 
\Big[
A_k^{(2,0)} a_{k-1}^{(0)} + B_k^{(2,0)} a_k^{(0)} +C_k^{(2,0)} a_{k+1}^{(0)} 
\Big]
\notag\\& +
\omega_2 
\Big[
A_k^{(0,2)} a_{k-1}^{(0)} + B_k^{(0,2)}a_k^{(0)} + C_k^{(0,2)} a_{k+1}^{(0)} 
\Big]
+ 
\sum_{m = 0}^{j-2} D_m
a_{k-m}^{(1)} =0.
\label{eq:recrelappendixb2nd}
\end{align}
We note that these equations correspond to the perturbative equations in Eqs~\eqref{eq:first} and \eqref{eq:k-th}.

First, at ${\cal O}(\alpha^0)$, 
we obtain $\omega_0$ using Leaver's continued fraction method
by setting a large integer $k_{\rm max}$.
Next, at ${\cal O}(\alpha^1)$, we solve the equation
\begin{align}
a_{k_{\rm max}}^{(1)} = 0,
\label{eq:akmax1st}
\end{align}
directly with respect to $\omega_1$.
For this purpose, we rewrite 
$a_{k_{\rm max}}^{(1)}$ as a function of $\omega_0, \omega_1, a_0^{(0)}, a_0^{(1)}$ by using
Eqs.~\eqref{eq:recrelappendixb0th}-\eqref{eq:recrelappendixb1st} recursively,
then $a_{k_{\rm max}}^{(1)}$ depends on $\omega_1$ linearly.
This implies that we obtain a unique $\omega_1$ if we fix the value of $\omega_0$.
In a similar way, we can solve the equation 
\begin{align}
a_{k_{\rm max}}^{(2)} = 0,
\label{eq:akmax2nd}
\end{align}
directly with respect to $\omega_2$.
In the calculation, we can set $a_0^{(0)} = a_0^{(1)} = a_0^{(2)} = 1$ without loss of generality.
We have confirmed that this method can reproduce a consistent result with Table~\ref{tab:diagonal}.
We finally note that we do not need to
perform the Gaussian elimination 
to obtain the three term recursion relations
at ${\cal O}(\alpha^1)$ and higher order analysis 
unlike usual Leaver's continued fraction method~\cite{leaver1985, konoplya2011},
and this is also one of the advantage of our perturbative approach.

\section{Outlook}\label{sec:outlook}
In this paper, we proposed a systematic way to compute high-order perturbative corrections to black hole quasinormal mode frequencies with continuous deformation parameters. Our method is widely applicable to many situations, and allows to compute the high-order corrections very accurately. 
We showed various explicit examples. In particular, for the Reissner-Nordstr\"om black holes, we can expand the quasinormal mode frequency not only around the chargeless limit but also around the extremal limit. 

There are several future directions. It is interesting to consider the near extremal expansion of the Kerr black holes.
It was argued that the QNM frequencies in the extremal Kerr geometry have an interesting behavior in \cite{Sasaki:1989ca}. 
It is also interesting to develop the perturbative expansion of rotating black holes in modified gravity theories~\cite{Srivastava:2021imr,Pierini:2021jxd,Pierini:2022eim,Wagle:2021tam,Li:2022pcy,Cano:2023tmv,Cano:2020cao,Cano:2023jbk,Cano:2021myl}. In this case, the full analytic solution with the general rotating parameter 
is not yet known.
We inevitably have to restrict ourselves to the perturbative treatment in terms of the rotating parameter.
We would like to extend our framework to coupled master equations. 
Typically, the master equations in general relativity are decoupled, but in modified gravity theories, they are sometimes coupled~\cite{Molina:2010fb,Sarbach:2001mc,Cardoso:2018ptl,McManus:2019ulj,Nomura:2021efi,Cano:2021myl,Hui:2022vov}. Therefore if we consider perturbative expansions of modified parameters, it is desirable to generalize our formalism to such a situation.

\acknowledgments{
This research is supported by JSPS KAKENHI Grant Nos. JP22K03641 (YH) and JP22K03626 (MK).
}

\appendix

\section{Recursion relations among coefficients in parameterized QNM approach}\label{app:ejidentities}
When the master equation is given in a series expansion of a small parameter,
there is an ambiguity of the effective potential due to the choise of the master variable.
In this appendix, we first give a general discussion of the ambiguity of effective potential
by extending the result in~\cite{kimura2020}. This ambiguity leads to recursion relations among coefficients in the parameterized QNM approach.

\subsection{Parameterized QNM approach}
We consider the case with $f = f_0 = 1- r_H/r$, and the master equation is given by
\begin{align}
f \frac{d}{dr}\left(f\frac{d\Phi}{dr}\right) + (\omega^2 - V)\Phi = 0,
\label{eq:mastereqparaqnm}
\end{align}
with
\begin{align}
V &= V_0 + \delta V,
\\
\delta V &= \frac{f}{r_H^2} \sum_{j = 0}^{\infty} \alpha_j \left(\frac{r_H}{r}\right)^j,
\label{eq:parameterizedpotential}
\end{align}
where 
$V_0$ is the effective potential for non-perturbative case and
$\alpha_j$ denote the small parameters which can be written as series of a single parameter $\alpha$
\begin{align}
\alpha_j = \sum_{i = 1}^\infty \alpha^i A^{(i)}_{j}.
\label{eq:parameterizedA}
\end{align}
We note that many systems can be written in this form of the master equation~\cite{cardoso2019, mcmanus2019, hatsuda2020b}.
The QNM frequency behaves
\begin{align}
\omega = \omega_0 + \sum_{j=0}^\infty \alpha_j e_j + \sum_{j,k =0}^\infty \alpha_j \alpha_k e_{j,k} + \cdots.
\label{eq:parameterizedomega}
\end{align}
where $e_j, e_{j,k}, \cdots$ are model independent coefficients in parameterized QNM approach.

When $V_0$ is the Regge-Wheeler potential for the odd parity gravitational perturbation,
the coefficients are related to the coefficients appearing in subsection \ref{subsec:para} as
\begin{align}
    &e_j = e_j^{(1)},\qquad e_{j,j} = e_j^{(2)},\\ 
    &e_{j, k}=e_{k,j}= \frac{e_{j,k}^{(1,1)}}{2}\quad (j < k),
\end{align}
where numerical values of $e_j^{(1)}, e_j^{(2)}, e_{j,k}^{(1,1)}$ can be seen in Tables~\ref{tab:diagonal} and \ref{tab:off-diagonal}.

\subsection{Ambiguity of effective potential}
In this subsection, we use the coordinate $x$ defined by $dx/dr = 1/f$.
The master equation Eq.~\eqref{eq:mastereqparaqnm} in this coordinate becomes
\begin{align}
\frac{d^2 \Phi}{dx^2} + (\omega^2 - V)\Phi = 0.
\end{align}
We introduce a new variable $\Psi$ as\footnote{
Note that signatures of $X$ and $Y$ are opposite from~\cite{kimura2020}.
}
\begin{align}
\Psi = \left(1 + X \right) \Phi + 
Y  \frac{d\Phi}{dx},
\end{align}
where $X$ and $Y$ are ${\cal O}(\alpha)$ functions of $x$.
If $X$ and $Y$ satisfy the relation
\begin{align}
-Y^2 \frac{dV}{dx} 
+ Y\left(2 (\omega^2 - V)\frac{dY}{dx} 
- \frac{d^2 X}{dx^2}\right)
+
(1 + X)\left(2 \delta \frac{dX}{dx} + \frac{d^2Y}{dx^2}\right) = 0,
\label{eq:schrodingercondition}
\end{align}
$\Psi$ satisfies an equation
\begin{align}
\frac{d^2 \Psi}{dx^2} + (\omega^2 - V - \delta W) \Psi = 0,
\end{align}
where $\delta W$ is given by\footnote{
$\delta W$ also can be written in the form
$\delta W = (2 dX/dx + d^2Y/dx^2)/Y$.
}
\begin{align}
\delta W 
&=
\frac{1}{1+X}\left(Y \frac{dV}{dx}
 - 2 (\omega^2 - V) \frac{dY}{dx} + \frac{d^2X}{dx^2}\right),
\label{eq:deltaW}
\end{align}
and this denotes the ambiguity of effective potential.
We can regard that the effective potential changes
\begin{align}
V \to V + \delta W
\end{align}
due to the change of the master variable,
and then the small parameters $\alpha_j$ in Eq~\eqref{eq:parameterizedpotential} are also changed.
Eq.~\eqref{eq:schrodingercondition} can be integrated as
\begin{align}
2 C + Y\left((V-\omega^2) Y + \frac{dX}{dx}\right) 
- \frac{dY}{dx} 
- 2X - X\left(X + \frac{dY}{dx}\right)
=0,
\label{eq:integratedschcond}
\end{align}
where $C$ is the constant of integration.
If we expand 
\begin{align}
X &= \sum_{i = 1}^\infty \alpha^i X_i,
\\
Y &= \sum_{i = 1}^\infty \alpha^i Y_i,
\\
V &= V_0 + \delta V = \sum_{i = 0}^\infty \alpha^i V_i,
\\
\omega^2 &= \sum_{i = 0}^\infty \alpha^i {\cal E}_i,
\\
C &= \sum_{i = 1}^\infty \alpha^i {\cal C}_i,
\end{align}
Eq.~\eqref{eq:integratedschcond} can be solved order by order as
\begin{align}
X_i &= C_i - \frac{1}{2}Y_i^\prime
-
\frac{1}{2}\sum_{k = 1}^{i-1} \sum_{j = 0}^{i-k-1} ({\cal E}_j - V_j) Y_k Y_{i-k-j}
+
\frac{1}{2}\sum_{k=1}^{i-1}
\left(
Y_{i-k} X_k^\prime
-
X_{i-k} Y_k^\prime
-
X_{i-k} X_k 
\right).
\label{eq:resultxi}
\end{align}
If we also expand $\omega = \sum_{i=0}^\infty \alpha^i \omega_i$, ${\cal E}_i$ is given by
\begin{align}
{\cal E}_i = \sum_{j = 0}^i \omega_{i-j}\omega_j.
\end{align}
Substituting the result~\eqref{eq:resultxi} into Eq.~\eqref{eq:deltaW}, 
we can calculate the deformation of the effective potential $\delta W$ as the series of $\alpha$
\begin{align}
\delta W &= \sum_{i = 1}^\infty \alpha^i W_i.
\label{eq:deltaW2}
\end{align}
From Eq~\eqref{eq:deltaW}, we can write $W_i$ as
\begin{align}
W_i = \frac{d^2X_i}{dx^2} + 
\sum_{j = 0}^{i-1}
\left(Y_{i-j}\frac{dV_{j}}{dx} - 2({\cal E}_j - V_j)\frac{dY_{i-j}}{dx}
\right)
-
\sum_{j = 1}^{i-1} W_{i-j}X_j.
\end{align}
For lower $i$, the explicit forms are
\begin{align}
X_1 &= C_1 - \frac{1}{2}\frac{dY_1}{dx},
\\
X_2 &= C_2 
-
\frac{1}{2}({\cal E}_0 - V_0) Y_1^2
+
\frac{1}{8}\left(\frac{dY_1}{dx}\right)^2
-
\frac{1}{2}
\left(C_1^2 +\frac{dY_2}{dx}\right)
-
\frac{1}{4} Y_1 \frac{d^2Y_1}{dx^2},
\end{align}
and
\begin{align}
W_1 &= 
 Y_1 \frac{dV_0}{dx}
-
2 ({\cal E}_0 - V_0)
\frac{dY_1}{dx}
-
\frac{1}{2}
\frac{d^3 Y_1}{dx^3},
\\
W_2 &= 
 (Y_2-C_1Y_1) \frac{dV_0}{dx}
-
2 ({\cal E}_0 - V_0)
\frac{d(Y_2-C_1Y_1)}{dx}
-
\frac{1}{2}
\frac{d^3 (Y_2-C_1Y_1)}{dx^3}
\\
&
+
\frac{Y_1}{2} \left(2 \frac{dV_1}{dx} + Y_1 \frac{d^2V_0}{dx^2}\right)
-
({\cal E}_0 - V_0)\left[
2\left(\frac{dY_1}{dx}\right)^2
+
 Y_1 \frac{d^2 Y_1}{dx^2}\right]
\\ & 
+\frac{dY_1}{dx}\left(
-2({\cal E}_1 - V_1)
+
\frac{5Y_1}{2} \frac{dV_0}{dx}
-
\frac{1}{2}\frac{d^3Y_1}{dx^3}
\right)
-
\frac{Y_1}{4}  \frac{d^4Y_1}{dx^2}.
\label{eq:W2}
\end{align}
We note that $W_i$ contains arbitrary functions $Y_1, Y_2, \cdots$.
If we set $V_i = 0$ for $i \ge 1$, the system is just a non-perturbative case whose effective potential is $V_0$.
Nevertheless, there is an ambiguity of effective potential due to the change of the master variable.
In this case, the ambiguity of effective potential does not change the QNM spectrum, 
and we can obtain recursion relations among coefficients in parameterized QNM approach by setting the functions $Y_i$ appropriately as shown in the next subsection.

\subsection{Recursion relations for odd parity case}
\subsubsection{Recursion relations from the Regge-Wheeler potential}
As an example, we consider the odd parity case
\begin{align}
V = V_0 = f_0\left(\frac{\ell(\ell + 1)}{r^2} - \frac{3 r_H}{r^3}\right).
\end{align}
In this case, ${\cal E}_1 = 0$ because there is no correction
term in the effective potential $V$, i.e., $V_i = 0$ for $i\ge 1$.
Setting\footnote{
From the degrees of freedom of $Y_2$, we can obtain the same relation as the first order relation among $e_j$.
Also, $C_1$ does not affect the result. Thus, we can set $Y_2 =0$ and $C_1 = 0$.
}
\begin{align}
Y_1 &= y_j \left(\frac{r_H}{r}\right)^j +  y_k \left(\frac{r_H}{r}\right)^k,
\\
Y_2 &= 0,
\\
C_1 &= 0,
\end{align}
where $j,k \ge -1$ are integers and $y_j, y_k$ are constants, Eqs~\eqref{eq:deltaW2}-\eqref{eq:W2} lead to
\begin{align}
& \delta V +\delta W  = 
\alpha y_j f_0 \left(\frac{r_H}{r}\right)^j
\bigg[
\frac{2 j {\cal E}_0}{r}
+\frac{(j+1) (j-2 \ell) (j+2 \ell+2)}{2
   r^3}
\notag\\&
-
\frac{(2 j+3) r_H \left(j (j+3)-2
   \left(\ell^2+\ell+3\right)\right)}{2
   r^4}
+\frac{(j-2) (j+2) (j+6) r_H^2}{2
   r^5}
\bigg] + (j \leftrightarrow k)
\notag\\& +
\alpha^2y_j^2 f_0 \left(\frac{r_H}{r}\right)^{2j}
\bigg[
-\frac{j
   (3 j+1) {\cal E}_0}{r^2}
+\frac{j (3 j+2)
   r_H {\cal E}_0}{r^3}
-\frac{3 (j+1)^2 (j-2 \ell) (j+2 \ell+2)}{4
   r^4}
\notag\\&
+\frac{(3
   j+4) r_H \left(3 j^3+12 j^2-j (8 \ell
   (\ell+1)+1)-2 (5 \ell (\ell+1)+9)\right)}{4
   r^5}
\notag\\&
-\frac{(3 j+5) r_H^2 \left(3 j^3+15 j^2-j
   (4 \ell (\ell+1)+7)-6
   \left(\ell^2+\ell+7\right)\right)}{4 r^6}
\notag\\&
+\frac{3 (j-2) (j+2)^2 (j+6)
   r_H^3}{4 r^7}
\bigg] +  (j \leftrightarrow k) 
\notag\\&
+
\alpha^2y_j y_k f_0 \left(\frac{r_H}{r}\right)^{j+k}
\bigg[
-\frac{{\cal E}_0 (j^2+4 j k+j+k^2+k)}{r^2}
+\frac{{\cal E}_0 r_H (j^2+j (4 k+2)+k (k+2))}{r^3}
\notag\\&
+\frac{1}{4 r^4}
\Big(
j^2 (-6 k+4 \ell (\ell+1)-11)-2 j^3 (k+3)-j^4
+4 (k (k+6)+6) \ell-k (k+1) (k+2) (k+3)
\notag\\&
+2 j (4 (2 k+3) \ell^2+4 (2 k+3) \ell-k (k (k+3)+4)-3)+4 (k (k+6)+6) \ell^2
\Big)
\notag\\&
+
\frac{r_H}{4 r^5}
\Big( j^2 (24 k-8 \ell (\ell+1)+47)+6 j^3 (k+4)+3 j^4
+k^2 (47-8 \ell (\ell+1))+3 k^4+24 k^3
\notag\\&
-2 k (31 \ell (\ell+1)+29)
-16 (5 \ell (\ell+1)+9)
+j (6 k^3+24 k^2-4 k (8 \ell (\ell+1)+1)-62 \ell (\ell+1)-58)
\Big)
\notag\\&
+
\frac{r_H^2}{4 r^6}
\Big(
j^2 (4 (\ell^2+\ell-17)-30 k)-6 j^3 (k+5)-3 j^4
+4 k^2 (\ell^2+\ell-17)
-3 k^4-30 k^3
\notag\\&
+k (38 \ell (\ell+1)+161)
+60 (\ell^2+\ell+7)
+j (-6 k^3-30 k^2+4 k (4 \ell (\ell+1)+7)+38 \ell (\ell+1)+161)\Big)
\notag\\&
+\frac{r_H^3 (2 j^3 (k+6)+4 j^2 (3 k+8)+j^4+2 j (k+6) (k^2-8)+k^2 (k+4) (k+8)-96 (k+3))}{4 r^7}
\bigg]
\notag\\&
+  {\cal O}(\alpha^3),
\label{eq:deltavdeltaw}
\end{align}
where we used the relation $d/dx = f d/dr$.
From this result, we can read $\alpha_i$ for $\delta V + \delta W$.
We decompose the coefficients $\alpha_i = A^{(1)}_i \alpha + A^{(2)}_i \alpha^2 + {\cal O}(\alpha^3)$ 
in Eq.~\eqref{eq:parameterizedA} as
\begin{align}
 A^{(1)}_i &= y_j \partial_{y_j} A^{(1)}_i  + y_k \partial_{y_k} A^{(1)}_i
\\
 A^{(2)}_i &= 
\frac{y_j^2}{2}\partial^2_{y_j} A^{(2)}_i  + 
\frac{y_k^2}{2}\partial^2_{y_k} A^{(2)}_i 
+y_jy_k \partial_{y_j}\partial_{y_k} A^{(2)}_i.
\end{align}
Introducing
$\partial_{y_j} A^{(1)}_i = r_H^{-1} B^{(1)}_i$,
$\partial_{y_j} \partial_{y_k} A^{(2)}_i = r_H^{-2} B^{(2)}_i$,
then one can see that the relations
\begin{align}
\partial_{y_k} A^{(1)}_i &=r_H^{-1} B^{(1)}_i|_{j\to k},
\\
\partial^2_{y_j} A^{(2)}_i &= \frac{r_H^{-2}}{2}B^{(2)}_i|_{k \to j},
\\
\partial^2_{y_k} A^{(2)}_i &= \frac{r_H^{-2}}{2}B^{(2)}_i|_{j \to k}
\end{align}
hold from the expression of Eq.~\eqref{eq:deltavdeltaw}.
The explicit forms of $B^{(1)}_i$ and $B^{(2)}_i$ become
\begin{align}
B^{(1)}_{j+1} &= 2  j r_H^2 {\cal E}_0,
\\
B^{(1)}_{j+3} &= \frac{1}{2}
(j+1) (j-2 \ell) (j+2 \ell+2),
\\
B^{(1)}_{j+4} &= -\frac{1}{2}
(2 j+3)  \left(j (j+3)-2
   \left(\ell^2+\ell+3\right)\right),
\\
B^{(1)}_{j+5} &= \frac{1}{2}
(j-2) (j+2) (j+6),
\end{align}
and 
\begin{align}
 B^{(2)}_{j+k+2} &= 
-(j^2+4 j k+j+k^2+k)r_H^2{\cal E}_0,
\\
 B^{(2)}_{j+k+3} &= (j^2+j (4 k+2)+k (k+2))
r_H^2{\cal E}_0, 
\\
 B^{(2)}_{j+k+4} &= 
\frac{1}{4}
\Big(
j^2 (-6 k+4 \ell (\ell+1)-11)-2 j^3 (k+3)-j^4
+4 (k (k+6)+6) \ell 
\notag\\&
-k (k+1) (k+2) (k+3)
+2 j (4 (2 k+3) \ell^2+4 (2 k+3) \ell
\notag\\&
-k (k (k+3)+4)-3)+4 (k (k+6)+6) \ell^2
\Big),
\\
 B^{(2)}_{j+k+5} &= 
\frac{1}{4}
\Big( j^2 (24 k-8 \ell (\ell+1)+47)+6 j^3 (k+4)+3 j^4
\notag\\&
+k^2 (47-8 \ell (\ell+1))+3 k^4+24 k^3
-2 k (31 \ell (\ell+1)+29)
-16 (5 \ell (\ell+1)+9)
\notag\\&
+j (6 k^3+24 k^2-4 k (8 \ell (\ell+1)+1)-62 \ell (\ell+1)-58)
\Big),
\\
 B^{(2)}_{j+k+6} &= 
\frac{1}{4}
\Big(
j^2 (4 (\ell^2+\ell-17)-30 k)-6 j^3 (k+5)-3 j^4
+4 k^2 (\ell^2+\ell-17)
\notag\\&
-3 k^4-30 k^3
+k (38 \ell (\ell+1)+161)
+60 (\ell^2+\ell+7)
\notag\\&
+j (-6 k^3-30 k^2+4 k (4 \ell (\ell+1)+7)+38 \ell (\ell+1)+161)\Big),
\\
 B^{(2)}_{j+k+7} &= 
\frac{1}{4}\Big(2 j^3 (k+6)+4 j^2 (3 k+8)+j^4+2 j (k+6) (k^2-8)
\notag\\&
+k^2 (k+4) (k+8)-96 (k+3)\Big).
\end{align}
Because ${\cal E}_1 = 0$ and then $\omega = \omega_0$, from Eq.~\eqref{eq:parameterizedomega}, we obtain a relation
\begin{align}
\sum_{j=0}^\infty \alpha_j e_j + \sum_{j,k =0}^\infty \alpha_j \alpha_k e_{j,k} = 0.
\label{eq:deltaomegaiszero}
\end{align}
From ${\cal O}(\alpha)$ and ${\cal O}(\alpha^2)$ terms in Eq.~\eqref{eq:deltaomegaiszero}, we obtain 
independent recursion relations among $e_j$ and $e_{j,k}$ 
\begin{align}
0 &= 
\sum_{a = 1}^{5} B_{j+a}^{(1)} e_{j+a}
\notag\\&=
 B_{j+1}^{(1)} e_{j+1} + B_{j+3}^{(1)} e_{j+3} + B_{j+4}^{(1)} e_{j+4} + B_{j+5}^{(1)} e_{j+5},
\label{eq:recrelation1st}
\end{align}
and
\begin{align}
0 &= 
\sum_{a,b = 1}^{5}  B_{j+a}^{(1)}B_{k+b}^{(1)}e_{j+a,k+b}
+
\frac{1}{2}\sum_{a = 2}^{7} B_{j+k+a}^{(2)} e_{j+k+a}
\notag\\
&=
B_{j+1}^{(1)}B_{k+1}^{(1)}e_{j+1,k+1}
+
B_{j+1}^{(1)}B_{k+3}^{(1)}e_{j+1,k+3}
+
B_{j+1}^{(1)}B_{k+4}^{(1)}e_{j+1,k+4}
+
B_{j+1}^{(1)}B_{k+5}^{(1)}e_{j+1,k+5}
\notag\\&
+
B_{j+3}^{(1)}B_{k+1}^{(1)}e_{j+3,k+1}
+
B_{j+3}^{(1)}B_{k+3}^{(1)}e_{j+3,k+3}
+
B_{j+3}^{(1)}B_{k+4}^{(1)}e_{j+3,k+4}
+
B_{j+3}^{(1)}B_{k+5}^{(1)}e_{j+3,k+5}
\notag\\&
+
B_{j+4}^{(1)}B_{k+1}^{(1)}e_{j+4,k+1}
+
B_{j+4}^{(1)}B_{k+3}^{(1)}e_{j+4,k+3}
+
B_{j+4}^{(1)}B_{k+4}^{(1)}e_{j+4,k+4}
+
B_{j+4}^{(1)}B_{k+5}^{(1)}e_{j+4,k+5}
\notag\\&
+
B_{j+5}^{(1)}B_{k+1}^{(1)}e_{j+5,k+1}
+
B_{j+5}^{(1)}B_{k+3}^{(1)}e_{j+5,k+3}
+
B_{j+5}^{(1)}B_{k+4}^{(1)}e_{j+5,k+4}
+
B_{j+5}^{(1)}B_{k+5}^{(1)}e_{j+5,k+5}
\notag\\&
+
\frac{1}{2}\Big[
B_{j+k+2}^{(2)}e_{j+k+2}
+
B_{j+k+3}^{(2)}e_{j+k+3}
+
B_{j+k+4}^{(2)}e_{j+k+4}
\notag\\&
+
B_{j+k+5}^{(2)}e_{j+k+5}
+
B_{j+k+6}^{(2)}e_{j+k+6}
+
B_{j+k+7}^{(2)}e_{j+k+7}\Big].
\label{eq:recrelation2nd}
\end{align}
We note again that $e_j = e_j^{(1)}, e_{j,j} = e_j^{(2)}$ and $e_{j,k} = e_{j,k}^{(1,1)}/2$ for $j \neq k$,
where numerical values of $e_j^{(1)}, e_j^{(2)}, e_{j,k}^{(1,1)}$ can be seen in Tables~\ref{tab:diagonal} and \ref{tab:off-diagonal}.
Using the first order recursion relation in Eq.~\eqref{eq:recrelation1st}, 
$e_j$ with higher $j$ can be written only from those with a few lower $j$, {\it i.e.,} $e_0, e_2$ and $e_7$~\cite{kimura2020}.
However, 
this is not the case for the second order recursion relation in Eq.~\eqref{eq:recrelation2nd}. 
In fact, to calculate $e_{j,k}$ with higher $j,k$ using Eq.~\eqref{eq:recrelation2nd},
we need the values of $e_{j,0}, e_{j,2}, e_{j,7}, e_{k,0}, e_{k,2}, e_{k,7}$.
To improve this point, 
we study the case with the potential which contains first order correction terms
in the next subsection.

\subsubsection{Improved recursion relation for $e_{j,k}$}
We consider the Regge-Wheeler potential with first order correction terms
\begin{align}
V &= V_0 + \delta V 
\notag\\&= f_0\left(\frac{\ell(\ell + 1)}{r^2} - \frac{3 r_H}{r^3}\right) 
+  \frac{\alpha f_0}{r_H^2}\left[v_{j}   \left(\frac{r_H}{r}\right)^{j+5} +  v_{k}  \left(\frac{r_H}{r}\right)^{k+5} \right],
\label{eq:potentialimproved1}
\end{align}
where $j,k \ge -1$ are integers and $v_{j}, v_{k}$ are constants.
We also assume that $j \neq 2$ and $k \neq 2$.
In this case, 
the QNM frequency behaves
\begin{align}
\omega = \omega_0 + \alpha \omega_1 + \alpha^2 \omega_2,
\label{eq:omegaappenb32no1}
\end{align}
with
\begin{align}
\omega_1 &= v_{j} e_{j+5} + v_{k} e_{k+5},
\label{eq:omegaappenb32no2}
\\
\omega_2 &= v_{j}^2 e_{j+5,j+5} + 2 v_{j} v_{k}e_{j+5,k+5} + v_{k}^2 e_{k+5,k+5}.
\label{eq:omegaappenb32no3}
\end{align}
${\cal E}_1 = 2 \omega_0 \omega_1$ becomes
\begin{align}
{\cal E}_1 = 2 v_{j} e_{j+5} \omega_0  + 2 v_{k} e_{k+5}\omega_0.
\end{align}
For this potential $V = V_0 + \delta V$, we set 
\begin{align}
Y_1 &= y_{j} \left(\frac{r_H}{r}\right)^j +  y_k \left(\frac{r_H}{r}\right)^k,
\\
Y_2 &= 0,
\\
C_1 &= 0,
\end{align}
with 
\begin{align}
y_j &= -\frac{2 v_j r_H}{(j-2)(j+2)(j+6)},
\\
y_k &= -\frac{2 v_k r_H}{(k-2)(k+2)(k+6)}.
\end{align}
Then, Eqs~\eqref{eq:deltaW2}-\eqref{eq:W2} lead to
\begin{align}
& \delta V +\delta W  = 
\alpha y_j f_0 \left(\frac{r_H}{r}\right)^j
\bigg[
\frac{2 j {\cal E}_0}{r}
+\frac{(j+1) (j-2 \ell) (j+2 \ell+2)}{2
   r^3}
\notag\\&
-
\frac{(2 j+3) r_H \left(j (j+3)-2
   \left(\ell^2+\ell+3\right)\right)}{2
   r^4}
\bigg] + (j \leftrightarrow k)
 +
\alpha^2 {\cal E}_1 \left[y_j f_0 \left(\frac{r_H}{r}\right)^j \frac{2j}{r}
+
y_k f_0 \left(\frac{r_H}{r}\right)^k \frac{2k}{r}
\right]
\notag\\& +
\alpha^2y_j^2 f_0 \left(\frac{r_H}{r}\right)^{2j}
\bigg[
-\frac{j
   (3 j+1) {\cal E}_0}{r^2}
+\frac{j (3 j+2)
   r_H {\cal E}_0}{r^3}
-\frac{3 (j+1)^2 (j-2 \ell) (j+2 \ell+2)}{4
   r^4}
\notag\\&
+\frac{(3
   j+4) r_H \left(3 j^3+12 j^2-j (8 \ell
   (\ell+1)+1)-2 (5 \ell (\ell+1)+9)\right)}{4
   r^5}
\notag\\&
-\frac{(3 j+5) r_H^2 \left(
j^3+3 j^2+j (1-4 \ell (\ell+1))-6 (\ell^2+\ell-1)
\right)}{4 r^6}
\notag\\&
-\frac{3 (j-2) (j+2)^2 (j+6)
   r_H^3}{4 r^7}
\bigg] +  (j \leftrightarrow k) 
\notag\\&
+
\alpha^2y_j y_k f_0 \left(\frac{r_H}{r}\right)^{j+k}
\bigg[
-\frac{{\cal E}_0 (j^2+4 j k+j+k^2+k)}{r^2}
+\frac{{\cal E}_0 r_H (j^2+j (4 k+2)+k (k+2))}{r^3}
\notag\\&
+\frac{1}{4 r^4}
\Big(
j^2 (-6 k+4 \ell (\ell+1)-11)-2 j^3 (k+3)-j^4
+4 (k (k+6)+6) \ell-k (k+1) (k+2) (k+3)
\notag\\&
+2 j (4 (2 k+3) \ell^2+4 (2 k+3) \ell-k (k (k+3)+4)-3)+4 (k (k+6)+6) \ell^2
\Big)
\notag\\&
+
\frac{r_H}{4 r^5}
\Big( j^2 (24 k-8 \ell (\ell+1)+47)+6 j^3 (k+4)+3 j^4
+k^2 (47-8 \ell (\ell+1))+3 k^4+24 k^3
\notag\\&
-2 k (31 \ell (\ell+1)+29)
-16 (5 \ell (\ell+1)+9)
+j (6 k^3+24 k^2-4 k (8 \ell (\ell+1)+1)-62 \ell (\ell+1)-58)
\Big)
\notag\\&
-
\frac{r_H^2}{4 r^6}
\Big(
2 j^2 (3 k-2 (\ell^2+\ell-4))
+2 j^3 (k+4)+j^4
+j (2 k^3+6 k^2-16 k \ell (\ell+1)+4 k-38 \ell (\ell+1)+23)
\notag\\&
-4 k^2 (\ell^2+\ell-4)+k^4+8 k^3+k (23-38 \ell (\ell+1))-60 (\ell^2+\ell-1)
\Big)
\notag\\&
-
\frac{r_H^3 (
2 j^3 (k+6)+4 j^2 (3 k+8)+j^4+2 j (k+6) (k^2-8)+k^2 (k+4) (k+8)-96 (k+3)
)}{4 r^7}
\bigg]
\notag\\&
+  {\cal O}(\alpha^3).
\label{eq:deltavdeltawno2}
\end{align}
We note that the above potential at ${\cal O}(\alpha)$
does not have terms with $(r_H/r)^{j+5}$ and $(r_H/r)^{k+5}$ unlike Eq.~\eqref{eq:deltavdeltaw}.
Similar to the discussion in the previous subsection,
we can read the coefficients $B_i^{(1)}$ and $B_i^{(2)}$ as
\begin{align}
B^{(1)}_{j+1} &= 2  j r_H^2 {\cal E}_0,
\\
B^{(1)}_{j+3} &= \frac{1}{2}
(j+1) (j-2 \ell) (j+2 \ell+2),
\\
B^{(1)}_{j+4} &= -\frac{1}{2}
(2 j+3)  \left(j (j+3)-2
   \left(\ell^2+\ell+3\right)\right),
\end{align}
and 
\begin{align}
 B^{(2)}_{j+1} &= -2 j (k-2)(k+2)(k+6) r_H^2 \omega_0 e_{k+5},
\\
 B^{(2)}_{k+1} &= -2 k (j-2)(j+2)(j+6) r_H^2 \omega_0 e_{j+5},
\\
 B^{(2)}_{j+k+2} &= 
-(j^2+4 j k+j+k^2+k)r_H^2{\cal E}_0,
\\
 B^{(2)}_{j+k+3} &= (j^2+j (4 k+2)+k (k+2))
r_H^2{\cal E}_0, 
\\
 B^{(2)}_{j+k+4} &= 
\frac{1}{4}
\Big(
j^2 (-6 k+4 \ell (\ell+1)-11)-2 j^3 (k+3)-j^4
+4 (k (k+6)+6) \ell 
\notag\\&
-k (k+1) (k+2) (k+3)
+2 j (4 (2 k+3) \ell^2+4 (2 k+3) \ell
\notag\\&
-k (k (k+3)+4)-3)+4 (k (k+6)+6) \ell^2
\Big),
\\
 B^{(2)}_{j+k+5} &= 
\frac{1}{4}
\Big( j^2 (24 k-8 \ell (\ell+1)+47)+6 j^3 (k+4)+3 j^4
\notag\\&
+k^2 (47-8 \ell (\ell+1))+3 k^4+24 k^3
-2 k (31 \ell (\ell+1)+29)
-16 (5 \ell (\ell+1)+9)
\notag\\&
+j (6 k^3+24 k^2-4 k (8 \ell (\ell+1)+1)-62 \ell (\ell+1)-58)
\Big),
\\
 B^{(2)}_{j+k+6} &= 
-\frac{1}{4}
\Big(
2 j^2 (3 k-2 (\ell^2+\ell-4))
+2 j^3 (k+4)+j^4 
\notag\\&
+j (2 k^3+6 k^2-16 k \ell (\ell+1)+4 k-38 \ell (\ell+1)+23)
\notag\\&
-4 k^2 (\ell^2+\ell-4)+k^4+8 k^3+k (23-38 \ell (\ell+1))-60 (\ell^2+\ell-1)
\Big),
\\
 B^{(2)}_{j+k+7} &= 
-\frac{1}{4}\Big((
2 j^3 (k+6)+4 j^2 (3 k+8)+j^4+2 j (k+6) (k^2-8)
\notag\\&
+k^2 (k+4) (k+8)-96 (k+3)
)\Big).
\end{align}
Then, the QNM frequency can be calculated from Eq.~\eqref{eq:parameterizedA},
and it should be same as Eq.~\eqref{eq:omegaappenb32no1} with Eqs~\eqref{eq:omegaappenb32no2} and~\eqref{eq:omegaappenb32no3}.
{}From this condition, we obtain independent recursion relations at ${\cal O}(\alpha^2)$ as
\begin{align}
& 
\frac{1}{2}(j-2)(j+2)(j+6)(k-2)(k+2)(k+6)
e_{j+5,k+5}
\notag\\&= 
2\sum_{a,b = 1}^{4}  B_{j+a}^{(1)}B_{k+b}^{(1)}e_{j+a,k+b}
+
\sum_{a = 2}^{7} B_{j+k+a}^{(2)} e_{j+k+a}
+
B_{j+1}^{(2)} e_{j+1}
+
B_{k+1}^{(2)} e_{k+1}.
\label{eq:recrelation2ndno2}
\end{align}
We note again that $j,k \ge -1$ and $j\neq 2, k \neq 2$ in the above equation.

In fact, we can obtain further independent recursion relations for $e_{j,k}$.
We consider the potential in Eq.~\eqref{eq:potentialimproved1} with 
$j \ge -1, j \neq 2$ and $k \ge -5$.
Setting
\begin{align}
Y_1 &= y_{j} \left(\frac{r_H}{r}\right)^j,
\\
Y_2 &= 0,
\\
C_1 &= 0,
\end{align}
with 
\begin{align}
y_j &= -\frac{2 v_j r_H}{(j-2)(j+2)(j+6)},
\end{align}
we can calculate $\delta V +\delta W$ 
from Eqs~\eqref{eq:deltaW2}-\eqref{eq:W2}, 
and derive the recursion relations similar to the above discussion.
Here, we only show the result:
\begin{align}
0 &= 
(j-2) (j+2) (j+6) e_{j+5,k+5}
-(2 j+3) (j (j+3)-2 (\ell^2+\ell+3)) e_{j+4,k+5}
\notag\\&
+(j+1) (j-2 \ell) (j+2 \ell+2) e_{j+3,k+5}
+4 j r_H^2 \mathcal{E}_0 e_{j+1,k+5}
\notag\\&
+4 j r_H^2 \omega_0 e_{j+1} e_{k+5}
-(2 j+k+5) e_{j+k+6}
+(2 j+k+6) e_{j+k+7}.
\label{eq:recrelation2ndno3}
\end{align}
Using Eqs.~\eqref{eq:recrelation2ndno2} and~\eqref{eq:recrelation2ndno3},
the second order coefficients $e_{j,k}$ with higher $j,k$ can be written by those with $j,k \le 7$
and the first order coefficients $e_j$.\footnote{
Some of coefficients $e_{j,k}$ with $j,k \le 7$ are not independent.
For example, we can choose
$e_{0,0}, e_{1,0}, e_{1,1}, e_{2,0}, e_{2,1}, e_{2,2}, e_{3,0}, e_{3,1}, e_{3,2}, e_{3,3}, 
e_{7,0}, e_{7,1}, e_{7,2}, e_{7,3}$ and $e_{7,7}$
as independent $e_{j,k}$, then, the other $e_{j,k}$ can be written by these.
}
We note that we can derive recursion relations for higher order $\alpha$ from a straightforward extension of the above 
discussion.

\subsection{Reduction of the effective potential}\label{app:potentialreduction}

Using the ambiguity of effective potential, we can reduce the 
the effective potential so that $\delta V$ only has lower order coefficients $\alpha_j$.
In~\cite{kimura2020}, the first order case is discussed, but in fact, the discussion holds even for the higher order case.
For the linear order case, 
we can reduce effective potential by using $O(\alpha)$ ambiguity according to~\cite{kimura2020}.
For the quadratic order case, setting $Y_1 = 0$ and $Y_2 = y_j (r_H/r)^j$
for the odd parity perturbation, the form of the ambiguity of the effective potential at $O(\alpha^2)$ 
becomes same as the linear order case. 
Then, from the same discussion as linear case in~\cite{kimura2020}, we can reduce the effective potential at $O(\alpha^2)$ so that $\delta V$ only has $\alpha_0, \alpha_1, \alpha_2$ and $\alpha_7$ terms. Repeating this process to higher order, we can reduce $O(\alpha^n)$ effective potential.

\bibliographystyle{amsmod}
\bibliography{QNM}

\end{document}